\documentclass{article}
\usepackage{graphicx} 
\usepackage{amsmath}
\usepackage{hyperref}
\hypersetup{colorlinks,linkcolor={blue},citecolor={blue},urlcolor={blue}}
\usepackage[version=4]{mhchem}
\usepackage{mathtools, nccmath}
\usepackage{setspace}
\usepackage{amssymb}
\usepackage{etoc}
\usepackage{cite}
\usepackage{float}
\usepackage{lipsum}
\usepackage{blindtext}
\usepackage[table,xcdraw]{xcolor}
\usepackage{multirow}
\usepackage{booktabs}
\usepackage{siunitx}
\usepackage{booktabs}
\usepackage{outline}
\usepackage{avant} 
\usepackage{multirow}
\usepackage[table,xcdraw]{xcolor}

\usepackage[a4paper, margin=1in, centering]{geometry}
\usepackage{array}
\usepackage{rotating}
\usepackage{epstopdf}
\usepackage{tikz}
\usepackage{lscape}
\usepackage{subcaption}
\usepackage[toc,page]{appendix}
\usepackage{enumitem}

\title{Embedding Generalized CP Symmetry in One Zero Texture Neutrino Mass Models}

\author{{Priya}{\footnote{kashyappriya963@gmail.com},  Simran Arora\footnote{009simranarora@gmail.com}, and B. C. Chauhan\footnote{bcawake@hpcu.ac.in}}}

\date{\textit{Department of Physics and Astronomical Science, Central University of Himachal Pradesh, Dharamshala (HP) 176215, India}}

\begin{document}
    \maketitle  
    \begin{abstract}
         
         In this study, we investigate one zero textures within the framework of generalized CP symmetry associated with the complex tribimaximal matrix. By combining these approaches, we derive predictive neutrino mass matrices and establish correlations between different parameters. Future long-baseline experiments, including NO$\nu$A, DUNE, and Hyper-Kamiokande, will provide a test of the predictions for the atmospheric mixing angle arising from this analysis. We also analyze the implications for neutrinoless double beta decay in the context of one-zero textures, in light of current and future experimental data. Our analysis indicates that the sum of the three neutrino masses in the inverted hierarchy is inconsistent with the cosmological bounds from Planck data ($\Sigma m_i < 0.12 \text{eV}$). In addition, constraints from the DESI/SDSS+Pantheon+DES-SN dataset ($\Sigma m_i < 0.17\text{eV}$) within the $\Lambda$CDM + Fluid DR + $\sum m_\nu$ framework at 95\% confidence level also disfavor the inverted hierarchy. In addition, constraints from the DESI experiment ($\Sigma m_i < 0.072$eV) may rule out all matrices in the $X_1$ case except $m_I$, and in the $X_2$ scenario except $m_I$, $m_{II}$, and $m_{III}$. Thus, the improved cosmological observations on $\Sigma m_i$ shall have decisive implications for viability of this class of model.

    \end{abstract}


\section{Introduction}

The Standard Model (SM) of particle physics is the theory that describes three of the four known fundamental forces in the universe (electromagnetic, weak, and strong interactions, excluding gravity) and classifies all known elementary particles with massless neutrinos. Its remarkable success strongly indicates that the SM will remain an excellent approximation to nature at distance scales as small as $10^{-18}$ m \cite{Gaillard:1998ui}. The signatures of solar and atmospheric neutrino oscillations detected by the Super-Kamioka Neutrino Detection Experiment (Super-Kamiokande) \cite{Super-Kamiokande:1998kpq}  and the Sudbury Neutrino Observatory (SNO)\cite{SNO:2002tuh} later validated by the Kamioka Liquid Scintillator Antineutrino Detector (KamLAND) experiment\cite{KamLAND:2002uet}, which confirmed that neutrinos possess a very small mass, thereby revealing the limitations of the SM. After measurement of the reactor mixing angle $\theta_{13}$ from the experiments Daya Bay \cite{Leitner:2017jco}, RENO \cite{RENO:2012mkc} and Double Chooz \cite{DoubleChooz:2011ymz}, all the three mixing angles ($\theta_{12}$, $\theta_{23}$, $\theta_{13}$), solar mass squared difference, i.e, $\Delta m_{\text{solar}}^2$ and magnitude of atmospheric mass squared difference, i.e, $\Delta m_{\text{atm}}^2$ have been measured successfully with considerable precision \cite{Ding:2013bpa}. The determination of neutrino mass hierarchy, matter-antimatter asymmetry, the nature of neutrinos, and CP violation in the lepton sector still remain unsolved mysteries. Therefore, these issues are the primary focus of upcoming experiments. The best fit values with errors($\sigma$'s) of mixing angles and mass squared differences are listed in Table \ref{Tab1}\cite{Esteban:2024eli}. 

\begin{table}[H]
  \centering
  \begin{tabular*}{\textwidth}{@{\extracolsep{\fill}} c c c c}
    \hline
    \hline
    S.No. & Neutrino Parameters & Best Fit ($1 \sigma$) & ($3 \sigma$) \\
    \hline
    1 & $\Delta m_{\text{solar}}^2$/$10^{-5}$ $eV^2$ & $7.49^{+0.19}_{-0.19}$ & 6.92 - 8.05 \\
    2 & $\Delta m_{\text{atm}}^2$/$10^{-3}$ $eV^2$ & $2.513^{+0.021}_{-0.019}$ & 2.451 - 2.578 \\
    3 & $\theta_{12}$/\(^\circ\) & $33.68^{+0.73}_{-0.70}$ & 31.63 - 35.95 \\
    4 & $\theta_{13}$/\(^\circ\) & $8.56^{+0.11}_{-0.11}$ & 8.19 - 8.89 \\
    5 & $\theta_{23}$/\(^\circ\) & $43.3^{+1.0}_{-0.8}$ & 41.3 - 49.9 \\
    
    \hline
    \hline
  \end{tabular*}
  \caption{Best fit values of neutrino parameters with $1 \sigma$ and $3 \sigma$ ranges taken from \cite{Esteban:2024eli}. }
  \label{Tab1}
\end{table}

The confirmation of nonzero neutrino masses from neutrino oscillation experiments is the first significant evidence of physics beyond the Standard Model \cite{Mohapatra:2005wg}. The discovery of $\nu_{\mu} \rightarrow \nu_{e}$ appearance by T2K in 2013~\cite{T2K:2013ppw}, later confirmed by the NuMI Off-Axis $\nu_{e}$ Appearance (NO$\nu$A) experiment~\cite{Smith:2022ezh}, holds significant importance providing the way for exploring three-flavor effects. In the framework of three active neutrinos, mass and flavor eigenstates are related by the 3$\times$ 3 mixing matrix, commonly referred as the Pontecorvo-Maki-Nakagawa-Sakata (PMNS) matrix. The lepton mixing matrix is defined by nine parameters: three mass eigenstates, three mixing angles, and three phases, i.e., one Dirac type CP phase and two Majorana type CP phases (the two additional phases present in the Majorana case do not affect oscillations) \cite{Giganti:2017fhf}. The PMNS matrix in the standard parametrization \cite{ParticleDataGroup:2018ovx} is
\small
\begin{equation}
    U_{PMNS} =
\begin{pmatrix}
\cos \theta_{12} \cos \theta_{13} & \sin \theta_{12} \cos \theta_{13} & \sin \theta_{13} e^{-i\delta} \\
- \cos \theta_{23} \sin \theta_{12} - \cos \theta_{12} \sin \theta_{13} \sin \theta_{23} e^{i\delta} & \cos \theta_{12} \cos \theta_{23} - \sin \theta_{12} \sin \theta_{13} \sin \theta_{23} e^{i\delta} & \cos \theta_{13} \sin \theta_{23} \\
\sin \theta_{12} \sin \theta_{23} - \cos \theta_{12} \sin \theta_{13} \cos \theta_{23} e^{i\delta} & - \cos \theta_{12} \sin \theta_{23} - \sin \theta_{12} \sin \theta_{13} \cos \theta_{23} e^{i\delta} & \cos \theta_{13} \cos \theta_{23}
\end{pmatrix} P.
\label{eqn1}
\end{equation}

Here, $\theta_{12}$, $\theta_{13}$, $\theta_{23}$ are the mixing angles, $\delta$ is the Dirac CP phase and matrix P is a diagonal matrix representing a phase matrix consisting two Majorana phases. The tribimaximal (TBM) mixing pattern is one of the most extensively studied lepton mixing schemes, often explored through the application of flavor symmetries\cite{He:2011gb,Plentinger:2005kx,Dev:2010pf,Ahn:2011ep}. A key prediction of the TBM mixing pattern is that the reactor mixing angle, \( \theta_{13} \), should be zero. Additionally, the atmospheric mixing angle is expected to be maximal, i.e., \( \theta_{23} = \frac{\pi}{4} \). Furthermore, TBM mixing predicts a specific value for the solar mixing angle, \( \theta_{12} = \sin^{-1}\left(\frac{1}{\sqrt{3}}\right) \) given as 
\begin{equation}
    U_{\text{TBM}} = \begin{pmatrix}
\frac{\sqrt{2}}{\sqrt{3}} & \frac{1}{\sqrt{3}} & 0 \\
-\frac{1}{\sqrt{6}} & \frac{1}{\sqrt{3}} & \frac{1}{\sqrt{2}} \\
\frac{1}{\sqrt{6}} & \frac{1}{\sqrt{3}} & -\frac{1}{\sqrt{2}}
\end{pmatrix}.
\label{eqn3}
\end{equation}

The Majorana phases will be disappeared if neutrinos are Dirac particles. We can express the lepton mass matrix as 
\begin{equation}
    m_\nu = U^{*} \, \text{diag}(m_1, m_2, m_3) \, U^{\dagger}.
\label{eqn4}
\end{equation}

The mixing matrix \( U \) can exhibit specific patterns due to residual flavor symmetries of the neutrino mass matrix \( m_\nu \) \cite{Samanta:2016wca}. These residual symmetries occur when \( m_\nu \) remains invariant under certain non-trivial transformations \( G_i \) \cite{Ding:2024ozt}

\begin{equation}
    G_i = U d_i U^{\dagger}, \quad i = 1, 2, 3, 4 ,
\end{equation}

where the \( d_i \) are given as
\begin{equation}
    d_1 = \text{diag}(1, -1, -1), \quad d_2 = \text{diag}(-1, 1, -1),  \quad d_3 = \text{diag}(-1, -1, 1), \quad d_4 = \text{diag}(1, 1, 1).
\label{eqn26}
\end{equation}

The exploration of residual flavor symmetries in the neutrino mass matrix has led to the study of other types of symmetries, including the generalized remnant CP transformation. Unlike residual flavor symmetries, which leave the neutrino mass matrix \( m_\nu \) unchanged, these generalized CP transformations convert \( m_\nu \) into its complex conjugate \( m_\nu^* \). These transformations are represented by \( X_i \) and are connected to the lepton mixing matrix \( U \) by the relation
\begin{equation}
    X_i = U d_i U^T \quad \text{for} \quad i = 1, 2, 3, 4.    
\label{eqn5}
\end{equation}

The imposition of any two of these CP symmetries results in modified TBM mixing patterns, each associated with a single residual flavor symmetry. Specifically, the action of two CP transformations, $X_i X_j^*$, is equivalent to a flavor symmetry transformation $G_k$, where $i$, $j$, and $k$ are cyclic permutations of $\{1, 2, 3\}$. This equivalence is crucial as it allows the lepton mixing matrix to be derived from residual CP symmetries \cite{Chen:2015nha}. Moreover, using generalized CP symmetries, rather than just flavor symmetries, constrains the Majorana CP phases, making this approach particularly effective as compared to the flavor symmetries.

In addition to residual flavor and remnant CP symmetries, alternative approaches such as imposing texture zeros\cite{Gautam:2018izb,Dev:2006qe,Singh:2016qcf,Singh:2022ijf, Verma:2020gpl, Verma:2021koo}, vanishing cofactors\cite{Dev:2015lya,Dev:2017jdx}, hybrid textures\cite{Dev:2013nua,Dev:2020xzq,Kalita:2015tda,Dev:2009he, Ankush:2021opd, Ankush:2023pax, Dev:2010vy}, or equalities\cite{Ismael:2021jay} among the neutrino mass matrix elements have also been explored to explain neutrino masses and mixings. Assuming that neutrinos are Majorana particles and the charged lepton mass matrix is diagonal, texture zeros in the neutrino mass matrix represent one of the simplest frameworks consistent with current and future neutrino experimental data. The combined use of flavor symmetry and texture zeros or vanishing cofactors in constructing neutrino mass matrix has led to highly predictive models. The another promising approach involves imposing texture zeros in conjunction with generalized CP symmetries. 

In this work, we aim to explore the implications of one zero textures within the framework of generalized CP symmetries $X_1$ and $X_2$. By combining these two approaches, we have constructed a more predictive model for the neutrino mass matrix. Specifically, we have examined patterns of one zero textures that are consistent with the symmetries imposed by $X_1$ and $X_2$. The presence of the complex tribimaximal matrix, $G_i \neq X_i$, allows us to distinguish between flavor symmetry and generalized CP symmetry. The preservation of flavor symmetries is thoroughly explained in \cite{Chen:2018zbq}. The $G_i = X_i$ case with texture zeros is examined in detail in \cite{Kumar:2023iaj}. That work uses three rotation angles $\theta_1$, $\theta_2$, and $\theta_3$ in the mixing matrix, while in our framework this structure is reduced to a single rotation angle. This change leads to different correlation patterns and modified predictions. In our analysis, we focus only on the $G_1$ and $G_2$ flavor symmetries, however $G_3$ corresponds to a reactor mixing angle equal to zero, which is experimentally excluded.

\maketitle

\section{Generalized CP symmetries}
In this section, we adopt the theoretical framework and analytical formalism developed in Refs.~\cite{Ding:2024ozt, Chen:2018zbq}. The mixing matrix \( U \) shows specific patterns due to residual flavor symmetries of the neutrino mass matrix \( m_\nu \). These residual symmetries occur when \( m_\nu \) remains invariant under certain non-trivial transformations \( G_i \), such that

\begin{equation}
    G_i^T m_\nu G_i = m_\nu \quad \text{for} \quad i = 1, 2, 3, 4.
\end{equation}

Here, \( G_4 \) is the trivial identity matrix, and we can check that
\[
G_i^2 = 1, \quad \text{and} \quad G_i G_j = G_j G_i = G_k \quad \text{with} \quad i \neq j \neq k \neq 4.
\]

The residual flavor symmetry of the Majorana neutrino mass matrix corresponds to the Klein group, isomorphic to $Z_2 \times Z_2$. Both quark and lepton mass matrices can exhibit simultaneous remnant flavor and CP symmetries. The explicit form of these symmetries is constrained by the experimentally observed lepton mixing matrix. In this section, we explore detailed structure of the remnant flavor and CP symmetries, examining their parametrization by combining both symmetries and how they contribute to the phenomenology of neutrino masses and mixing.

For the Majorana neutrinos, the most general form of the corresponding neutrino mass matrix in the charged lepton diagonal basis is 

\begin{equation}
    m_{\nu}^{\text{TBM}} = U_{\text{TBM}} \, \text{diag}(m_1, m_2, m_3) \,U_{TBM}^T , 
\label{a2}
\end{equation}
such that 
\begin{equation}
    m_{\nu}^{\text{TBM}} = \frac{1}{6} 
\begin{pmatrix}
4m_1 + 2m_2 & -2m_1 + 2m_2 & -2m_1 + 2m_2 \\
-2m_1 + 2m_2 & m_1 + 2m_2 + 3m_3 & m_1 + 2m_2 - 3m_3 \\
-2m_1 + 2m_2 & m_1 + 2m_2 - 3m_3 & m_1 + 2m_2 + 3m_3
\end{pmatrix}.
\label{a1}
\end{equation}

The neutrino mass matrix represented by Eqn. \ref{a1} remains invariant under the following residual flavor transformations given as

\begin{equation}
    \begin{array}{ccc}
G^{\text{TBM}}_1 = \frac{1}{3} \begin{pmatrix}
1 & -2 & -2 \\
-2 & -2 & 1 \\
-2 & 1 & -2
\end{pmatrix}, &
G^{\text{TBM}}_2 = \frac{1}{3} \begin{pmatrix}
-1 & 2 & 2 \\
2 & -1 & 2 \\
2 & 2 & -1
\end{pmatrix}, &
G^{\text{TBM}}_3 = - \begin{pmatrix}
1 & 0 & 0 \\
0 & 0 & 1 \\
0 & 1 & 0
\end{pmatrix}.
\end{array}
\label{a3}
\end{equation}

In other words, the neutrino mass matrix \( m_{\nu}^{\text{TBM}} \) satisfies
\[
(G^{\text{TBM}}_i)^T m_{\nu}^{\text{TBM}} G^{\text{TBM}}_i = m_{\nu}^{\text{TBM}},
\]
for \( i = 1, 2, 3 \).

\subsection*{Complex TBM Matrix}

Here, we explore into the complex tribimaximal(cTBM) matrix, incorporating non-zero Majorana phases\cite{Ding:2024ozt, Chen:2018zbq,Chen:2018eou}. Such that the cTBM matrix is given as
\begin{equation}
    U_{cTBM} = 
    \begin{pmatrix}
    \frac{\sqrt{2}}{\sqrt{3}} & \frac{e^{-i\rho}}{\sqrt{3}} & 0 \\
    -\frac{e^{i\rho}}{\sqrt{6}} & \frac{1}{\sqrt{3}} & \frac{e^{-i\sigma}}{\sqrt{2}} \\
    \frac{e^{i(\rho+\sigma)}}{\sqrt{6}} & -\frac{e^{i\sigma}}{\sqrt{3}} & \frac{1}{\sqrt{2}} \\
    \end{pmatrix}.
\label{eqn6}
\end{equation}

This cTBM mixing matrix predicts the same mixing angles as the usual real TBM pattern in Eqn. \ref{eqn3}, though the Majorana phases are non-vanishing. The exploration of residual flavor symmetries in the neutrino mass matrix \( m_\nu \) has extended to include generalized CP symmetries, which transform the mass matrix into its complex conjugate \( m_\nu^* \). These transformations are represented by \( X_i \) and are connected to the lepton mixing matrix \( U \) by the relation presented in Eqn. \ref{eqn5}. These symmetries are often studied in the context of finite discrete groups like \( \Delta(6n^2) \) \cite{Chen:2014wxa}, which naturally lead to trimaximal mixing patterns. This approach can yield specific predictions for the Dirac and Majorana CP phases, influencing ongoing experimental searches for neutrino oscillation parameters and neutrinoless double-beta decay (\( 0\nu\beta\beta \)) observables \cite{Chen:2015nha}. The generalized CP symmetry \( X_i \) is present in \( m_\nu \) if 
\begin{equation}
    X_i^T m_\nu X_i = m_\nu^*.
\end{equation}

The four CP symmetry matrices $X_{1,2,3,4}$ associated with the cTBM mixing pattern are
\begin{equation}
    X_{i} = U_{cTBM}d_{i}U^T_{cTBM},
\label{q2}
\end{equation}

where \( d_{1,2,3,4} \) are diagonal matrices as given in Eqn. \ref{eqn26}. Thus, the four CP symmetries are given in matrix form as

\begin{equation}
\begin{aligned}
    X_1 &= \frac{1}{6}
    \begin{pmatrix}
        4 - 2e^{-2i\rho} & -2e^{-i\rho} - 2e^{i\rho} & 2e^{i(\rho + \sigma)} + 2e^{-i(\rho + \sigma)} \\
        -2e^{-i\rho} - 2e^{i\rho} & -2 + e^{2i\rho} - 3e^{-2i\sigma} & -3e^{-i\sigma} - e^{i(2\rho + \sigma)} + 2e^{i\sigma} \\
        2e^{i(\rho + \sigma)} + 2e^{-i(\rho + \sigma)} & -3e^{-i\sigma} - e^{i(2\rho + \sigma)} + 2e^{i\sigma} & -3 + e^{2i(\rho + \sigma)} - 2e^{2i\sigma}
    \end{pmatrix}, \\
    X_2 &= \frac{1}{6}
    \begin{pmatrix}
        -4 + 2e^{-2i\rho} & 2e^{-i\rho} + 2e^{i\rho} & -2e^{i(\rho + \sigma)} - 2e^{-i(\rho + \sigma)} \\
        2e^{-i\rho} + 2e^{i\rho} & 2 - e^{2i\rho} - 3e^{-2i\sigma} & -3e^{-i\sigma} + e^{i(2\rho + \sigma)} - 2e^{i\sigma} \\
        -2e^{i(\rho + \sigma)} - 2e^{-i(\rho + \sigma)} & -3e^{-i\sigma} + e^{i(2\rho + \sigma)} - 2e^{i\sigma} & -3 - e^{2i(\rho + \sigma)} + 2e^{2i\sigma}
    \end{pmatrix}, \\
    X_3 &= \frac{1}{6}
    \begin{pmatrix}
        -4 - 2e^{-2i\rho} & -2e^{-i\rho} + 2e^{i\rho} & -2e^{i(\rho + \sigma)} + 2e^{-i(\rho + \sigma)} \\
        -2e^{-i\rho} + 2e^{i\rho} & -2 - e^{2i\rho} + 3e^{-2i\sigma} & 3e^{-i\sigma} + e^{i(2\rho + \sigma)} + 2e^{i\sigma} \\
        -2e^{i(\rho + \sigma)} + 2e^{-i(\rho + \sigma)} & 3e^{-i\sigma} + e^{i(2\rho + \sigma)} + 2e^{i\sigma} & 3 - e^{2i(\rho + \sigma)} - 2e^{2i\sigma}
    \end{pmatrix}, \\
    X_4 &= \frac{1}{6}
    \begin{pmatrix}
        4 + 2e^{-2i\rho} & 2e^{-i\rho} - 2e^{i\rho} & 2e^{i(\rho + \sigma)} - 2e^{-i(\rho + \sigma)} \\
        2e^{-i\rho} - 2e^{i\rho} & 2 + e^{2i\rho} + 3e^{-2i\sigma} & 3e^{-i\sigma} - e^{i(2\rho + \sigma)} - 2e^{i\sigma} \\
        2e^{i(\rho + \sigma)} - 2e^{-i(\rho + \sigma)} & 3e^{-i\sigma} - e^{i(2\rho + \sigma)} - 2e^{i\sigma} & 3 + e^{2i(\rho + \sigma)} + 2e^{2i\sigma}
    \end{pmatrix}.
\end{aligned}
\label{eqn5.4}
\end{equation}

The CP symmetries corresponding to the ``standard" real TBM matrix of Eqn. \ref{eqn3} are obtained by taking the limit \( \rho, \sigma \to 0 \). These CP symmetries are therefore given as

\[
X_1 = \frac{1}{3}
\begin{pmatrix}
1 & -2 & 2 \\
-2 & -2 & -1 \\
2 & -1 & -2
\end{pmatrix}, \quad
X_2 = \frac{1}{3}
\begin{pmatrix}
-1 & 2 & -2 \\
2 & -1 & -2 \\
-2 & -2 & -1
\end{pmatrix},
\]

\[
X_3 =
\begin{pmatrix}
-1 & 0 & 0 \\
0 & 0 & 1 \\
0 & 1 & 0
\end{pmatrix}, \quad
X_4 =
\begin{pmatrix}
1 & 0 & 0 \\
0 & 1 & 0 \\
0 & 0 & 1
\end{pmatrix}.
\]

The residual flavor symmetries can be generated by the CP transformations as follows
\begin{equation}
\begin{aligned}
    G_1 &= X_2 X_3^* = X_3 X_2^* = X_4 X_1^* = X_1 X_4^*, \\
    G_2 &= X_1 X_3^* = X_3 X_1^* = X_4 X_2^* = X_2 X_4^*, \\
    G_3 &= X_1 X_2^* = X_2 X_1^* = X_4 X_3^* = X_3 X_4^*, \\
    G_4 &= X_1 X_1^* = X_2 X_2^* = X_3 X_3^* = X_4 X_4^*.
\end{aligned}
\label{eqn61}
\end{equation}

Notice that only three of the four CP and flavor symmetries are really independent. If any three of the four CP symmetries in
Eqn. \ref{eqn5.4} are imposed simultaneously, the neutrino mixing matrix would reduce to the cTBM matrix in Eqn. \ref{eqn6} with $\theta_{13}$ = 0. Therefore, we will impose only two or only one of these CP symmetries, so that realistic mixing patterns with non-vanishing $\theta_{13}$ and CP violation are obtained.

The lepton mixing patterns, such as tribimaximal, golden ratio, and bi-maximal mixing, are excluded by current neutrino oscillation data, particularly by the precise measurement of the reactor angle, $\theta_{13}$. These patterns must be revamped to align with experimental results and to provide meaningful theoretical predictions for CP violation~\cite{Ding:2024ozt, Chen:2018zbq}.

We assume neutrinos are Majorana particles and demonstrate how the application of residual flavor and CP symmetries can lead to systematic generalizations of the mixing patterns. By imposing these residual symmetries, we can determine the $i$-th column of the mixing matrix, thereby deriving generalized patterns that are both viable and predictive. Consequently, the mixing matrix can be characterized by a limited set of parameters.

\subsection*{ Case I : $G_1$ flavor and $X_1,X_4$ CP symmetries}

As stated above, applying the generalized CP transformations represented by $X_1$ and $X_4$, the neutrino mass matrix $m_\nu$ transforms into its complex conjugate. The CP transformations $X_1$ and $X_4$ act as symmetries of $m_\nu$, which results in the preservation of the $G_1$ flavor symmetry. Consequently, the neutrino mass matrix $m_\nu$ must satisfy the symmetry constraints imposed by both the CP transformations $X_1$ and $X_4$ as~\cite{Ding:2024ozt, Chen:2018zbq}

\begin{equation}
    X^T_1m_{\nu}X_1=m^*_\nu, \hspace{1cm}        X^T_4m_{\nu}X_4=m^*_\nu.
\label{eqn7}
\end{equation}

Therefore, the light neutrino mass matrix is of the following form
\begin{equation}
     m'_{\nu} = U^T_{cTBM}m_{\nu}U_{cTBM} = \begin{pmatrix}
m_1 & 0 & 0 \\
0 & m_2 & \delta m \\
0 & \delta m & m_3 \\
\end{pmatrix},
\end{equation}

where the parameters $m_1, m_2, m_3$ and $\delta$m are real. The mass matrix $m'_\nu$ can be diagonalized by a real orthogonal matrix $R_{23}(\theta)$ given by

\begin{equation}
    R_{23}(\theta) = 
\begin{pmatrix}
1 & 0 & 0 \\
0 & \cos\theta & \sin\theta \\
0 & -\sin\theta & \cos\theta \\
\end{pmatrix},
\text{with}
\tan 2\theta = \frac{2\delta m}{m_3 - m_2}.
\end{equation}

As a result, in this case the lepton mixing matrix is given as

\begin{equation}
    U = U_{cTBM}R_{23}Q_\nu,
\end{equation}

such that

\begin{equation}
   U = \frac{1}{\sqrt{6}}
    \begin{pmatrix}
    2 & \sqrt{2}e^{-i\rho}\cos\theta & \sqrt{2}e^{-i\rho}\sin\theta \\
    -e^{i\rho} & \sqrt{2}\cos\theta - \sqrt{3}e^{-i\sigma}\sin\theta & \sqrt{2}\sin\theta + \sqrt{3}e^{-i\sigma}\cos\theta \\
    e^{i(\rho + \sigma)} & -\sqrt{3}\sin\theta - \sqrt{2}e^{i\sigma}\cos\theta & \sqrt{3}\cos\theta - \sqrt{2}e^{i\sigma}\sin\theta
    \end{pmatrix}
    Q_\nu.
\label{eqn8}
\end{equation}

Here $Q_\nu = \text{diag}(e^{ik_1\pi/2}, e^{ik_2\pi/2}, e^{ik_3\pi/2})$ is a diagonal unitary matrix with $k_{1,2,3} = 0, 1, 2, 3$ and $\theta$ is the rotation angle. The entries $\pm1$ represents the CP parities of the neutrino states and ensure that the neutrino mass eigenvalues remain non-negative. One can notice that, the first column of the lepton mixing matrix in Eqn. \ref{eqn8} is given by \(\frac{1}{\sqrt{6}} \begin{pmatrix} 2 \\ -e^{i \rho} \\ e^{i (\rho + \sigma)} \end{pmatrix}\). This column is consistent with that of the cTBM mixing pattern. It arises from the preserved $G_1$ symmetry. When other two CP symmetries \( X_2 \) and \( X_3 \) are imposed, the neutrino mass matrix also preserves the flavor symmetry \( G_1 = X_2 X_3^* = X_3 X_2^* \) as shown in Eqn \ref{eqn61}. We can calculate the neutrino mixing angles from a given mixing matrix \( U \) by using the following relations
\begin{equation}
    s^2_{12} = \frac{|U_{12}|^2}{1 - |U_{13}|^2}, \quad s^2_{23} = \frac{|U_{23}|^2}{1 - |U_{13}|^2}, \quad \text{and} \quad s^2_{13} = |U_{13}|^2.
\label{eqn121}
\end{equation}

The Jarlskog rephasing invariant (\(J_{CP}\)) is defined as 
\begin{equation}
    J_{CP} = \text{Im}(U_{11} U_{12}^* U_{21} U_{22}^*).
\end{equation}

\subsection*{ Case II : $G_2$ flavor and $X_2, X_4$ CP symmetries}

Accordingly, applying the generalized CP transformations represented by $X_2$ and $X_4$, the neutrino mass matrix $m_\nu$ transforms into its complex conjugate. The CP transformations $X_2$ and $X_4$ act as symmetries of $m_\nu$, which results in the preservation of the $G_2$ flavor symmetry. Consequently, the neutrino mass matrix $m_\nu$ must satisfy the symmetry constraints imposed by both the CP transformations $X_2$ and $X_4$ as~\cite{Ding:2024ozt, Chen:2018zbq}

\begin{equation}
    X^T_2m_{\nu}X_2=m^*_\nu,   \hspace{1cm}       X^T_4m_{\nu}X_4=m^*_\nu.
\label{eqn9}
\end{equation}

The light neutrino mass matrix will be 
\begin{equation}
     m'_{\nu} = U^T_{cTBM}m_{\nu}U_{cTBM} = \begin{pmatrix}
m_1 & 0 &  \delta m \\
0 & m_2 & 0\\
 \delta m & 0 & m_3 \\
\end{pmatrix},
\end{equation}

where the parameters $m_1, m_2, m_3$ and $\delta$m are real. The mass matrix $m'_\nu$ can be diagonalized by a real
orthogonal matrix $R_{13}(\theta)$ given by

\begin{equation}
    R_{13}(\theta) = 
\begin{pmatrix}
\cos\theta & 0 & \sin\theta \\
0 & 1 & 0 \\
-\sin\theta & 0 & \cos\theta \\
\end{pmatrix},
\text{with}
\tan 2\theta = \frac{2\delta m}{m_3 - m_1}.
\end{equation}

As a result, in this case the lepton mixing matrix is given as

\begin{equation}
    U = U_{cTBM}R_{13}Q_\nu,
\label{eqn44}
\end{equation}

\begin{equation}
   U = 
\frac{1}{\sqrt{6}}
\begin{pmatrix}
2\cos\theta & \sqrt{2}e^{-i\rho} & 2\sin\theta \\
-e^{i\rho}\cos\theta - \sqrt{3}e^{-i\sigma}\sin\theta & \sqrt{2} & -e^{i\rho}\sin\theta + \sqrt{3}e^{-i\sigma}\cos\theta \\
e^{i(\rho+\sigma)}\cos\theta - \sqrt{3}\sin\theta & -\sqrt{2}e^{i\sigma} & e^{i(\rho+\sigma)}\sin\theta + \sqrt{3}\cos\theta
\end{pmatrix}
Q_\nu.
\label{eqn10}
\end{equation}

One can notice that the second column of the mixing matrix is given by \(\frac{1}{\sqrt{3}} \begin{pmatrix} e^{-i \rho} \\ 1 \\ -e^{i \sigma} \end{pmatrix}\), which corresponds to the cTBM mixing pattern. This correspondence arises due to the preserved \( G_2 \) symmetry. When other two CP symmetries \( X_1 \) and \( X_3 \) are imposed, the neutrino mass matrix also preserves the flavor symmetry \( G_2 = X_1 X_3^* = X_3 X_1^* \) as shown in Eqn \ref{eqn61}. The \( G_3 \) symmetry is excluded from our analysis due to its prediction of a zero reactor mixing angle, which is experimentally disfavored. Ref.~[39] studies the case $G_i = X_i$, where the generalized CP transformation is aligned with the residual flavor symmetry. In contrast, our work focuses on the case $G_i \neq X_i$, this leads to different phase relations and modified constraints on the mass matrix entries. The expression for the atmospheric mixing angle($Sin^2\theta_{23}$) in $X_1$ and $X_2$ cases are summarized in the table \ref{tab:GiXi_1} and \ref{tab:GiXi} as
\begin{table}[H]
\centering
\begin{tabular}{c c}
\hline\hline
\textbf{$G_i = X_i$} & \textbf{$G_i \neq X_i$} \\
\hline
$ s_{23}^2 = \frac{1}{2}\left( 1 + \frac{\sqrt{6}\sin 2\theta}{3 - 2\sin^2\theta} \right) $
&
$ s_{23}^2 = \frac{1}{2} - \frac{\sqrt{6}\sin 2\theta\cos(\sigma)}{2\cos^2\theta + 4} $
\\ 
\hline\hline
\end{tabular}
\caption{Classification based on the relation between $G_i$ and $X_i$ for $X_1$ case.}
\label{tab:GiXi_1}
\end{table}
\begin{table}[H]
\centering
\begin{tabular}{c c}
\hline\hline
\textbf{$G_i = X_i$} & \textbf{$G_i \neq X_i$} \\
\hline
$ s_{23}^2 = \frac{1}{2}\left( 1 + \frac{\sqrt{3}\sin 2\theta}{3 - 2\sin^2\theta} \right) $
&
$ s_{23}^2 = \frac{1}{2} - \frac{\sqrt{3}\sin 2\theta\cos(\rho+\sigma)}{4\cos^2\theta + 2} $
\\ 
\hline\hline
\end{tabular}
\caption{Classification based on the relation between $G_i$ and $X_i$ for $X_2$ case.}
\label{tab:GiXi}
\end{table}

\section{Texture one zero with Generalized CP Symmetry}

The neutrino mass matrix remains invariant under residual symmetries and is transformed into its complex conjugate when a generalized CP transformation is applied. By imposing one zero textures in the neutrino mass matrix, we obtain six distinct patterns that are consistent with current neutrino oscillation data. These six patterns can be simultaneously diagonalized according to the conditions outlined in Eqn. \ref{eqn44}. Moreover, all six patterns satisfy the generalized CP conditions specified in Eqn. \ref{eqn7} and \ref{eqn9}. Here, we will explore the one zero textures within the framework of generalized CP symmetries

\[
\begin{array}{cc}
m_{I} = 
\begin{pmatrix}
0 & \times & \times \\
\times & \times & \times \\
\times & \times & \times \\
\end{pmatrix}, &
m_{II} = 
\begin{pmatrix}
\times & 0 & \times \\
0 & \times & \times \\
\times & \times & \times \\
\end{pmatrix}, \\
m_{III} = 
\begin{pmatrix}
\times & \times & 0 \\
\times & \times & \times \\
0 & \times & \times \\
\end{pmatrix}, &
m_{IV} = 
\begin{pmatrix}
\times & \times & \times \\
\times & 0 & \times \\
\times & \times & \times \\
\end{pmatrix}, \\
m_{V} = 
\begin{pmatrix}
\times & \times & \times \\
\times & \times & 0 \\
\times & 0 & \times \\
\end{pmatrix}, &
m_{VI} = 
\begin{pmatrix}
\times & \times & \times \\
\times & \times & \times \\
\times & \times & 0 \\
\end{pmatrix}.
\end{array}
\]

The presence of one zero textures in the neutrino mass matrix leads to a complex equation, which is given as
\begin{equation}
m_1 A + m_2 B + m_3 C = 0, 
\end{equation}
where \( A = U_{a1} U_{b1} \), \( B = U_{a2} U_{b2} \), \( C = U_{a3} U_{b3} \) and $U_{a1}, U_{b1}, U_{a2}, U_{b2}, U_{a3}, U_{b3}$ are the elements of mixing matrix. The above complex equation yields two mass ratios
\begin{equation}
\frac{m_1}{m_2} = \frac{\text{Re}(C)\text{Im}(B) - \text{Re}(B)\text{Im}(C)}{\text{Re}(A)\text{Im}(C) - \text{Re}(C)\text{Im}(A)}, \tag{33}
\end{equation}
and
\begin{equation}
\frac{m_1}{m_3} = \frac{\text{Re}(C)\text{Im}(B) - \text{Re}(B)\text{Im}(C)}{\text{Re}(B)\text{Im}(A) - \text{Re}(A)\text{Im}(B)}.
\end{equation}
Here \(\text{Re}\) (\(\text{Im}\)) denotes the real (imaginary) part. These mass ratios can be used to obtain the expression for the parameter \(R_\nu\), which is the ratio of mass squared differences \((\Delta m^2_{ij} = m_i^2 - m_j^2)\),
\begin{equation}
R_\nu \equiv \frac{\Delta m^2_{21}}{|\Delta m^2_{31}|} = \frac{\left(\frac{m_2}{m_1}\right)^2 - 1}{\left|\left(\frac{m_3}{m_1}\right)^2 - 1\right|}.
\label{qq}
\end{equation}
Given \(m_1 > m_3\) for an inverted mass hierarchy (IH) and \(m_1 < m_3\) for the normal mass hierarchy (NH).

\section{Results and Discussions}

The neutrino mass matrix is constructed using the mixing matrix given in Eqn. \ref{eqn8} and \ref{eqn10}, and subjected to generalized CP conditions derived from Eqn. \ref{q2}. We observed that all neutrino mass matrices with one zero textures are consistent with these generalized CP conditions, i.e., with \(X_1\) and \(X_2\). In the analysis, we have sampled $10^8$ points to scan the complete parameter space and used the mass squared differences \(\Delta m_{\text{solar}}^2\) and \(\Delta m_{\text{atm}}^2\), varying them within their 3\(\sigma\) ranges, as indicated in Table \ref{Tab1}. The rotation angle \(\theta\) varied from \(0\) to $2\pi$. Additionally, the Majorana phases ($\rho$ and $\sigma$) are explored freely over their entire permissible ranges i.e., from (0 - 2$\pi$) . The one zero textures constraint is applied by the parameter \(R_\nu\), representing mass ratios, which are allowed to lie within the 3\(\sigma\) range according to Eqn. \ref{qq}. Using the $3\sigma$ intervals of $\Delta m_{21}^2$ and 
$|\Delta m_{3l}^2|$ from NuFIT~6.0, the allowed range is
$
R_\nu \in [0.0265, 0.0328]$ for NH and $ R_\nu \in [0.0271, 0.0332]$ for IH. For each of the six texture one zero patterns, we computed the possible values of the effective Majorana mass for both normal hierarchy (NH) and inverted hierarchy (IH) with \(X_1\) and \(X_2\) respectively, as detailed in Tables \ref{Tab2} and \ref{tab3}. To assess the compatibility of the effective Majorana mass ($m_{ee}$), we incorporated constraints from the KamlandZen \cite{Gando:2020cxo,Shirai:2017jyz}, LEGEND \cite{Lopez-Castano:2019tgx}, CUORE \cite{CUORE:2022aiq}, and nEXO experiments \cite{nEXO:2021ujk}, which are represented by the horizontal lines in the figures. The cosmological limit on the sum of neutrino masses, $\Sigma m_i < 0.12\,\text{eV}$, provided by the Planck data \cite{Planck:2018vyg}, is shown as vertical lines in the figures where $m_{ee}$ is plotted against the sum of neutrino masses. The DESI experiment proposed a very stringent bound on sum of neutrino mass, i.e., 0.072eV. We also adopted the upper bound of $\sum m_\nu < 0.17 \, \text{eV}$, obtained from the combination of the P20$_H$+DESI/SDSS+DES-SN datasets in the $\Lambda$CDM + Fluid DR + $\sum m_\nu$ model at 95\% CL which provides the relaxation on Planck data\cite{Allali:2024aiv}.

\subsection*{Analysis of $X_1$ Condition}

For the $X_1$ condition, we obtained values for the \( m_{ee}\) as a function of the lightest neutrino mass constrained by the $R_\nu$ parameter at 3 $\sigma$ range, which are consistent with current and future experimental data, as detailed in Table \ref{Tab2}.

\begin{table}[H]
  \centering
  \small
  \begin{tabular*}{\textwidth}{@{\extracolsep{\fill}} c c c c c c}
    \hline
    \hline
    S.No. & Textures & Hierarchy & Effective Majorana mass(eV) & Lightest neutrino mass(eV) & Sum of mass(eV) \\
    \hline
    1 & {$m_I$} & Normal & 0.0 & 0.0019-0.0080 & 0.052-0.072\\
            &                        & Inverted & - & -&-\\
    \hline
    2 & \multirow{2}{*}{$m_{II}$} & Normal & 0.0050-0.29 & 0.023-0.24 & 0.10-0.73\\
      &                        & Inverted & 0.024-0.22 & 0.070-0.22 &0.21-0.68\\
    \hline
    3 & \multirow{2}{*}{$m_{III}$} & Normal & 0.0083-0.26 & 0.030-0.24 & 0.12-0.74\\
      &                        & Inverted & 0.024-0.23 & 0.070-0.24 & 0.19-0.72\\
    \hline
    4 & \multirow{2}{*}{$m_{IV}$} & Normal & 0.012-0.17 & 0.048-0.19 & 0.16- 0.60\\
      &                        & Inverted & 0.019-0.15&0.050-0.31 &0.19-0.94\\
    \hline
    5 & \multirow{2}{*}{$m_V$} & Normal & 0.011-0.10 & 0.051-0.14 & 0.17-0.44 \\
      &                        & Inverted & 0.058-0.15& 0.061-0.26& 0.21-0.86\\
    \hline
    6 & \multirow{2}{*}{$m_{VI}$} & Normal & 0.014-0.17 &0.049-0.17 & 0.16-0.52 \\
      &                        & Inverted & 0.019-0.18  & 0.050-0.21  & 0.19-0.65\\
    \hline
    \hline
  \end{tabular*}
  \caption{Values of the effective Majorana mass, lightest neutrino mass, and the sum of the three neutrino masses for both NH and IH for the X$_1$ condition.}

\label{Tab2}
\end{table}

The cosmological bound on the sum of neutrino masses, \(\Sigma m_i < 0.12\,\text{eV}\), excludes compatibility for the mass matrices $m_{IV}$, $m_{V}$, and $m_{VI}$ in NH. The inverted hierarchy (IH) is excluded from the analysis because all one zero texture matrices in the IH are inconsistent with Planck data and P20$_H$ + DESI/SDSS + DES-SN datasets, within the $\Lambda$CDM + \textit{Fluid DR} + $\sum m_\nu$ model at the 95\% CL.  The detailed description of the textures is given below:

\begin{enumerate}[label=\Roman*.]
\normalsize
    \item The mass matrix \( m_I \), which has a zero in the (1,1) position, cannot support inverted hierarchy because it predicts a large value for the reactor mixing angle (\( \theta_{13} \)), which is experimentally disfavored. The imposition of generalized CP conditions does not change this property. For the inverted hierarchy, the neutrino masses satisfy $m_3 < m_1 < m_2$, with typical values $m_1 \approx \sqrt{|\Delta m^2_{\rm atm}|} \sim 0.049~\rm eV$, $m_2 \sim 0.050~\rm eV$, and $m_3 \ll m_1, m_2$. The texture zero condition($m_I = 0$) for the $X_1$ case is
$$
\frac{3}{2}m_1 \;+\; 3\,e^{-2i\rho}\big(m_2\cos^2\theta + m_3\sin^2\theta\big)=0.
$$
Since the phase $\rho$ is unrestricted, satisfying the condition by tuning $e^{-2i\rho}$ would require fine adjustment.  
For a conservative estimate, we set $e^{-2i\rho}=1$ and neglect the small $m_3$ term
$$
\frac{3}{2}m_1 + 3m_2\cos^2\theta \approx 0.
$$
This gives
$$
\cos^2\theta = -\frac{m_1}{2m_2}.
$$
Substituting the inverted–hierarchy values,
$$
\cos^2\theta = -\frac{0.049}{2\times 0.050} \approx -0.49,
$$
which has no real solution.
This means that the neglected $m_3\sin^2\theta$ term must provide the entire negative contribution.  
Because $m_3$ is much smaller than $m_1$ and $m_2$, this forces $\sin^2\theta$ to become large, close to $1$.  
Using the mixing–matrix relation 
$$
\sin^2\theta_{13} = \frac{1}{3}\sin^2\theta,
$$
a large $\sin^2\theta$ implies
$$
\sin^2 2\theta_{13} \gtrsim 0.5,
$$
which is much higher than the experimental value ($\sim 0.085$). Therefore, this texture naturally drives $\theta_{13}$ to a large value in the inverted hierarchy, and the condition cannot be satisfied without phase tuning or without taking $m_3$ comparable to $m_1$ and $m_2$. Similar results have been shown in (\cite{Kumar:2023iaj}). For NH, the effective Majorana mass also vanishes. 
   
    \item Figure \ref{fig1}(\subref{fig1:a}) depicts the correlation between the \( m_{ee}\) and \(m_{\text{lightest}}\) for the mass matrix \(m_{II}\) in NH, characterized by a zero texture at the (1,2) position. The analysis reveals that this matrix is consistent with both current and future experimental constraints. The lower bound on \( m_{ee} \leq 0.0043 \, \text{eV} \) is within the sensitivity reach of KamlandZen, LEGEND, and nEXO experiment. The appearance of branches in the correlation plot is a direct consequence of imposing the \(R_{\nu}\) condition within the \(3\sigma\) range of the NuFIT\,6.0 oscillation data. This requirement restricts the allowed values of the Majorana phases. Under these constraints, the allowed solutions occur only along the boundary of the effective Majorana mass, while the intermediate region is excluded because the corresponding phase combinations do not satisfy the \(R_{\nu}\) condition at the \(3\sigma\) level. Figure \ref{fig1}(\subref{fig1:b}) illustrates the correlation between the \( m_{ee}\) and the sum of neutrino masses(\(\Sigma m_i\)). The results indicate compatibility with the P20H+DESI/SDSS+DES-SN dataset, \(\Sigma m_i < 0.17\,\text{eV}\) and also with cosmological bound.

\begin{figure}[H]  
    \centering
    
    \begin{subfigure}[b]{0.48\textwidth}
        \centering
        \includegraphics[width=\textwidth]{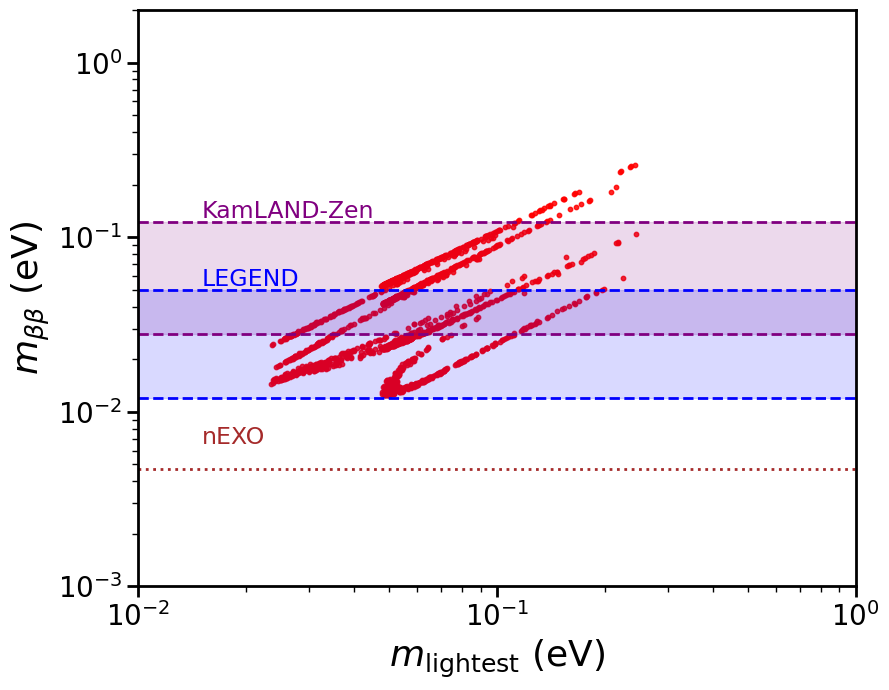}  
        \caption{}
        \label{fig1:a}
    \end{subfigure}\hfill
    \begin{subfigure}[b]{0.48\textwidth}
        \centering
        \includegraphics[width=\textwidth]{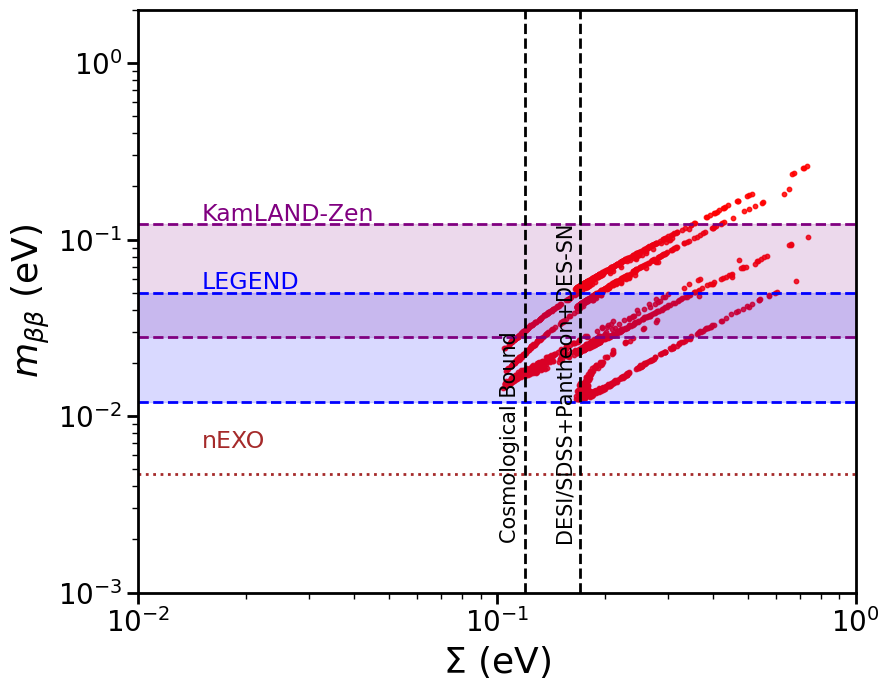}  
        \caption{}
        \label{fig1:b}
    \end{subfigure}
    
\caption{Effective Majorana mass with respect to lightest neutrino mass (a) and sum of neutrino mass (b) for mass matrix $m_{II}$ in NH for $X_1$.}
\label{fig1}
\end{figure}

\begin{figure}[H]  
    \centering
    
    \begin{subfigure}[b]{0.48\textwidth}
        \centering
        \includegraphics[width=\textwidth]{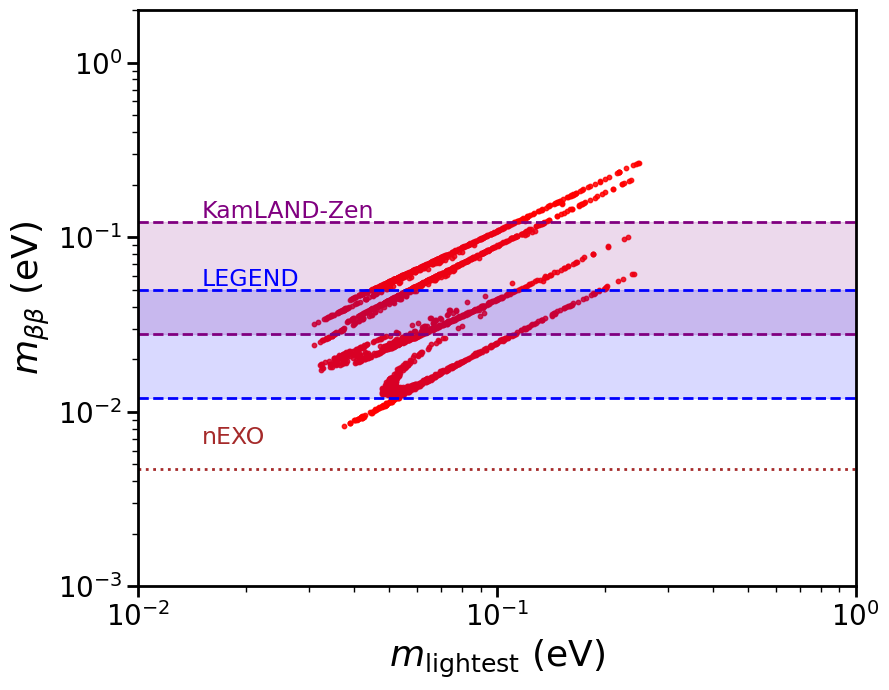}  
        \caption{}
        \label{fig2:a}
    \end{subfigure}\hfill
    \begin{subfigure}[b]{0.48\textwidth}
        \centering
        \includegraphics[width=\textwidth]{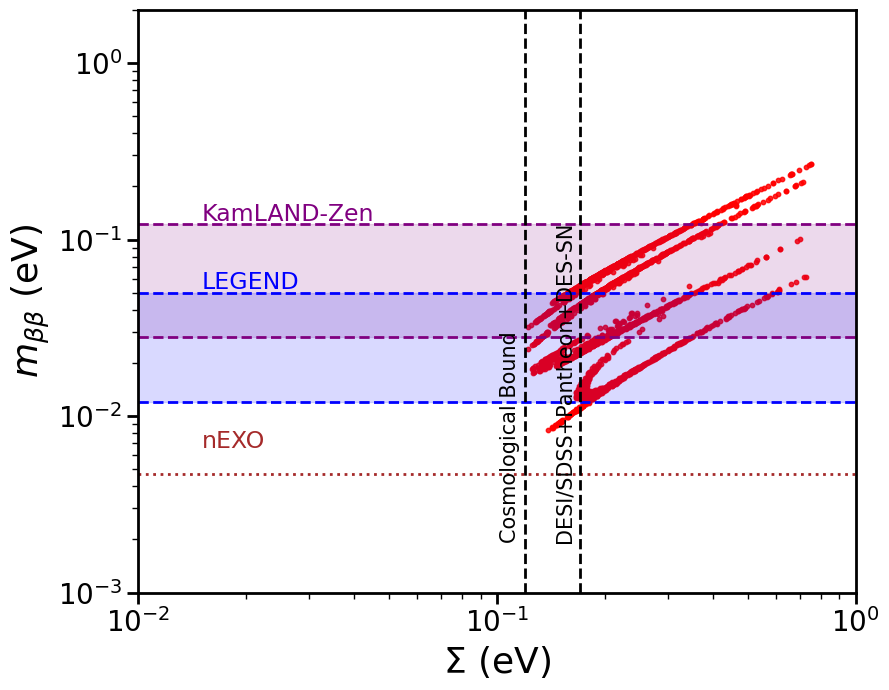}  
        \caption{}
        \label{fig2:b}
    \end{subfigure}
    
\caption{Effective Majorana mass with respect to lightest neutrino mass (a) and sum of neutrino mass (b) for mass matrix $m_{III}$ in NH for $X_1$.}
\label{fig2}
\end{figure}

\begin{figure}[H]  
    \centering
    
    \begin{subfigure}[b]{0.48\textwidth}
        \centering
        \includegraphics[width=\textwidth]{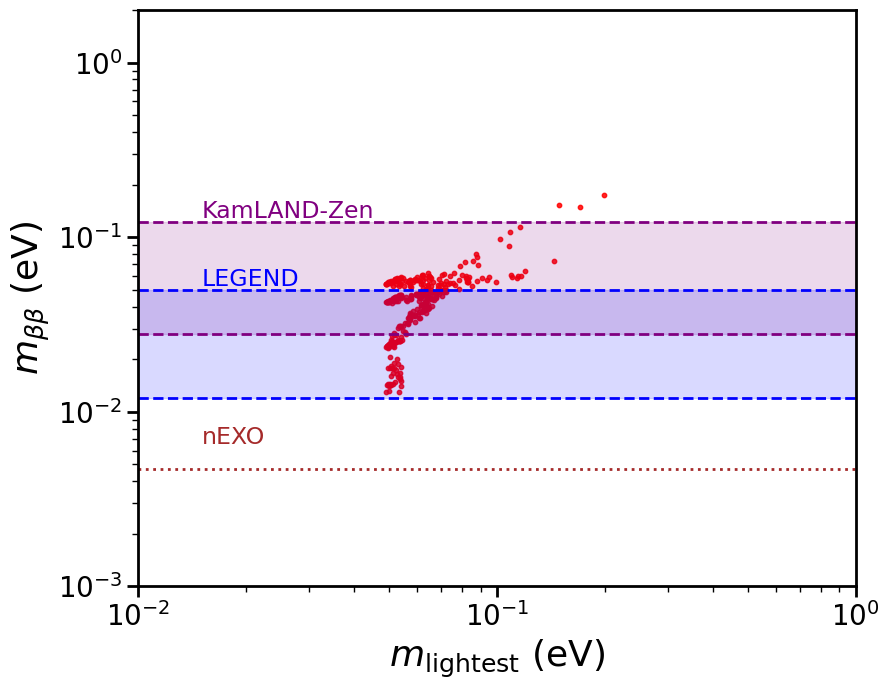}  
        \caption{}
        \label{fig3:a}
    \end{subfigure}\hfill
    \begin{subfigure}[b]{0.48\textwidth}
        \centering
        \includegraphics[width=\textwidth]{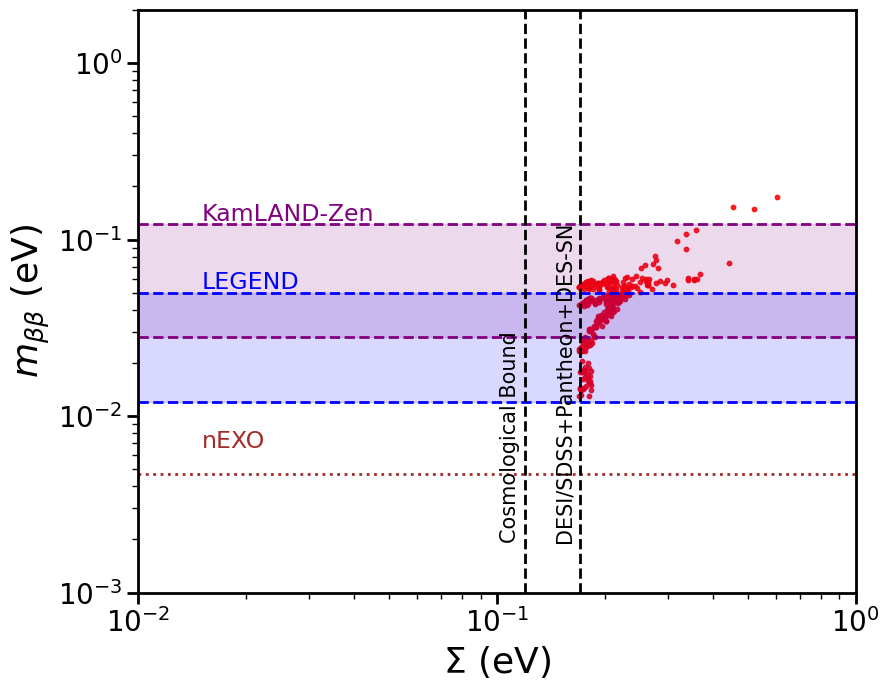}  
        \caption{}
        \label{fig3:b}
    \end{subfigure}
    
\caption{Effective Majorana mass with respect to lightest neutrino mass (a) and sum of neutrino mass (b) for mass matrix $m_{IV}$ in NH for $X_1$.}
\label{fig3}
\end{figure}

\begin{figure}[H]  
    \centering
    
    \begin{subfigure}[b]{0.48\textwidth}
        \centering
        \includegraphics[width=\textwidth]{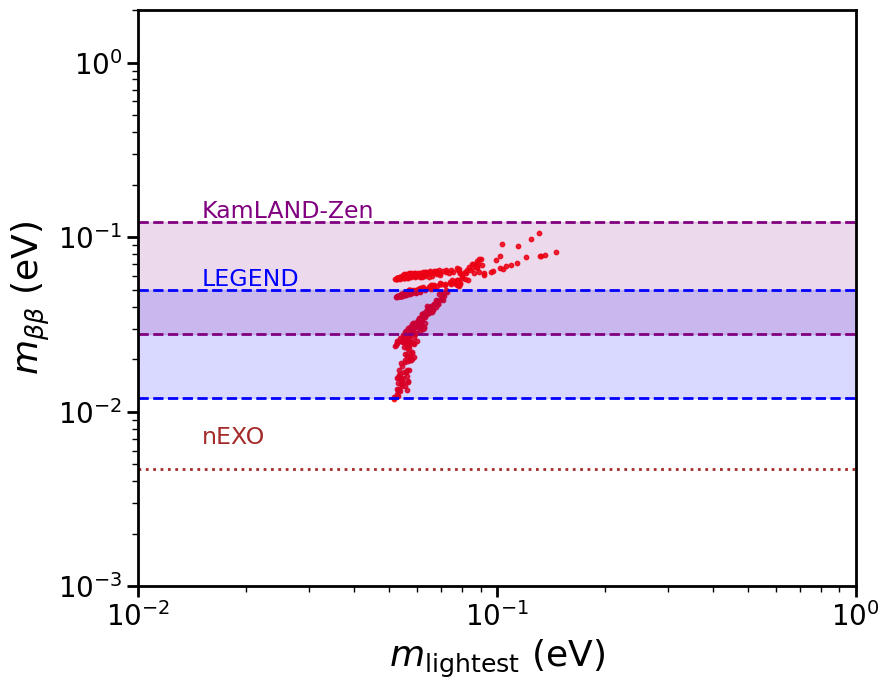}  
        \caption{}
        \label{fig4:a}
    \end{subfigure}\hfill
    \begin{subfigure}[b]{0.48\textwidth}
        \centering
        \includegraphics[width=\textwidth]{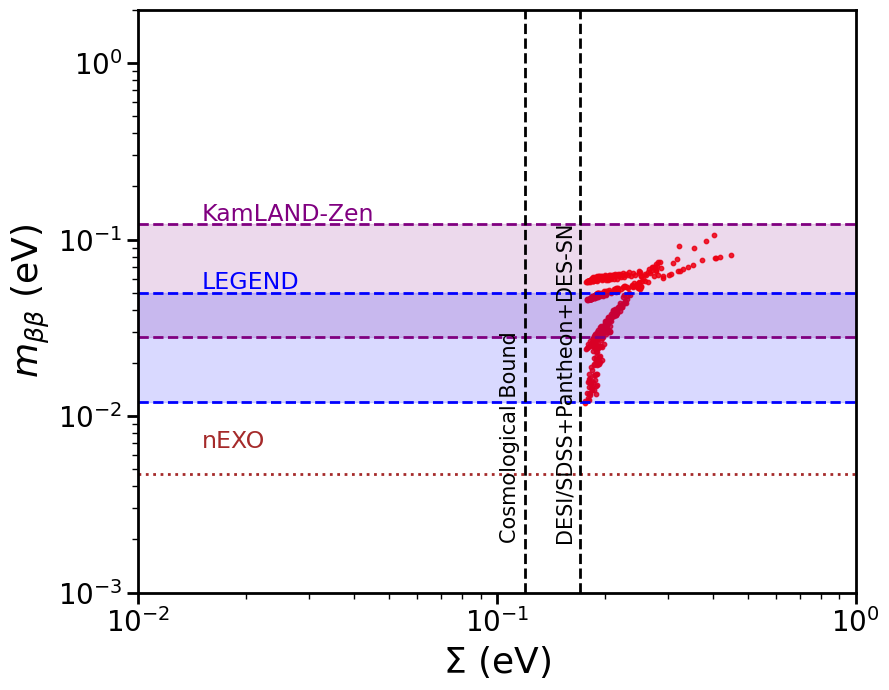}  
        \caption{}
        \label{fig4:b}
    \end{subfigure}
    
\caption{Effective Majorana mass with respect to lightest neutrino mass (a) and sum of neutrino mass (b) for mass matrix $m_{V}$ in NH for $X_1$.}
\label{fig4}
\end{figure}

\begin{figure}[H]  
    \centering
    
    \begin{subfigure}[b]{0.48\textwidth}
        \centering
        \includegraphics[width=\textwidth]{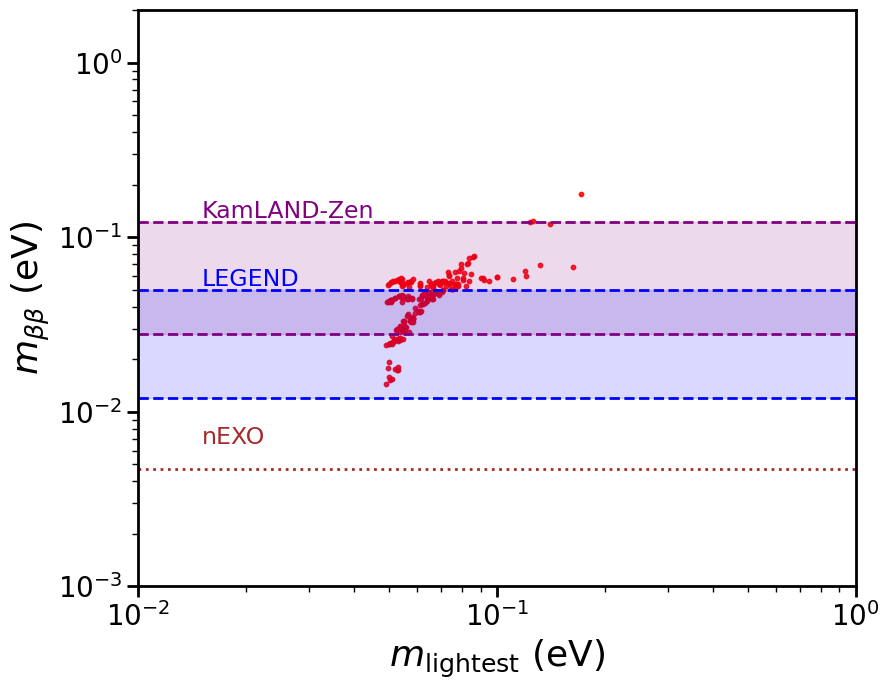}  
        \caption{}
        \label{fig5:a}
    \end{subfigure}\hfill
    \begin{subfigure}[b]{0.48\textwidth}
        \centering
        \includegraphics[width=\textwidth]{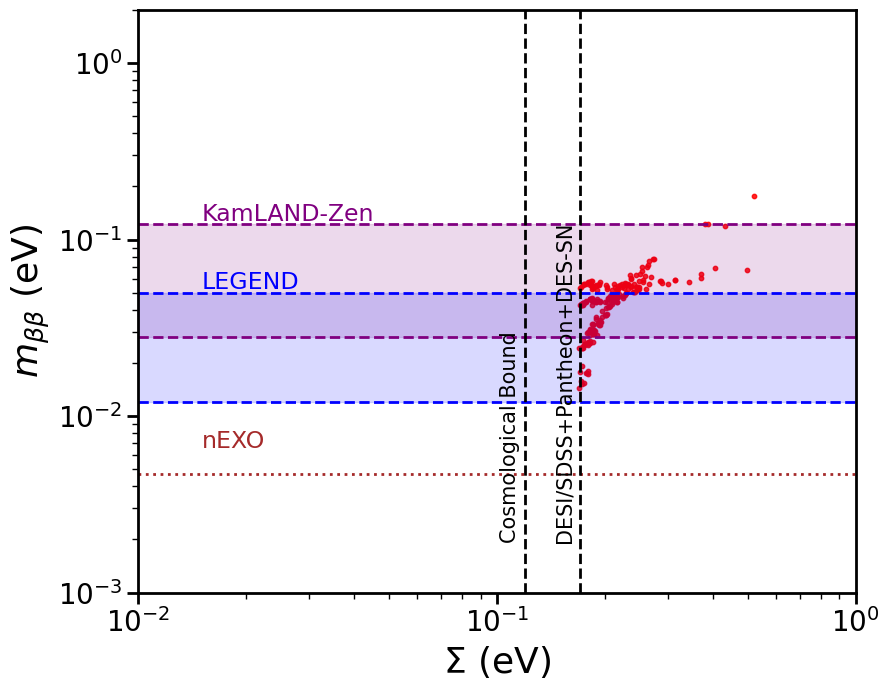}  
        \caption{}
        \label{fig5:b}
    \end{subfigure}
    
\caption{Effective Majorana mass with respect to lightest neutrino mass (a) and sum of neutrino mass (b) for mass matrix $m_{VI}$ in NH for $X_1$.}
\label{fig5}
\end{figure}

    \item Figure \ref{fig2}(\subref{fig2:a}) illustrates the correlation between \( m_{ee} \) and \( m_{\text{lightest}} \) for the mass matrix \( m_{III} \) in NH, having zero at (1,3) position. The lower bound on \( m_{ee} \leq 0.0022 \, \text{eV} \) is within the sensitivity reach of KamlandZen, LEGEND, and nEXO. Figure \ref{fig2}(\subref{fig2:b}) depicts the correlation between \( m_{ee} \) and \( \Sigma m_i \). The results indicate compatibility with the P20H+DESI/SDSS+DES-SN dataset, which constrains \( \Sigma m_i < 0.17\,\text{eV} \) and with cosmological bound.

    \item Figure \ref{fig3}(\subref{fig3:a}) shows the correlation between the \( m_{ee} \) and \( m_{\text{lightest}} \) for the mass matrix \( m_{IV} \) in the normal hierarchy (NH), with a zero at (2,2) position. The analysis demonstrates that this matrix is consistent with both current and future experimental constraints. The lower bound on \( m_{ee} \leq 0.0045 \, \text{eV} \) is within the sensitivity reach of KamlandZen and LEGEND experiment. Figure \ref{fig3}(\subref{fig3:b}) presents the correlation between \( m_{ee} \) and \( \Sigma m_i \). The results show that the matrix is compatible with P20H+DESI/SDSS+DES-SN dataset, but incompatible with the Planck limit.

    \item Figure \ref{fig4}(\subref{fig4:a}) depicts the correlation between the \( m_{ee}\) and \(m_{\text{lightest}}\) for the mass matrix \(m_{V}\) in NH, characterized by texture zero at the (2,3) position. The analysis reveals that this matrix is consistent with both current and future experimental constraints. The lower bound on \( m_{ee} \leq 0.0031\, \text{eV} \) is within the sensitivity reach of KamlandZen and LEGEND experiment. Figure \ref{fig4}(\subref{fig4:b}) illustrates the correlation between the \( m_{ee}\) and \(\Sigma m_i\). The results indicate incompatibility with the cosmological bound and also in tension with the P20H+DESI/SDSS+DES-SN dataset.
    
  \item Figure \ref{fig5}(\subref{fig5:a}) shows the correlation between the \( m_{ee}\) and \(m_{\text{lightest}}\) for the mass matrix \(m_{VI}\) in NH, characterized by a texture zero at the (3,3) position. The analysis reveals that this matrix is consistent with both current and future experimental constraints. The effective Majorana mass spans the range (0.0076 - 0.20)eV, sensitive to all the considered experimental range. Figure \ref{fig5}(\subref{fig5:b}) illustrates the correlation between the \( m_{ee}\) and \(\Sigma m_i\). The results indicate incompatibility with the cosmological bound, but compatible with the P20H+DESI/SDSS+DES-SN dataset on the sum of neutrino masses.

\end{enumerate}

\begin{figure}[H]  
    \centering
    
    \begin{subfigure}[b]{0.45\textwidth}
        \centering
        \includegraphics[width=\textwidth]{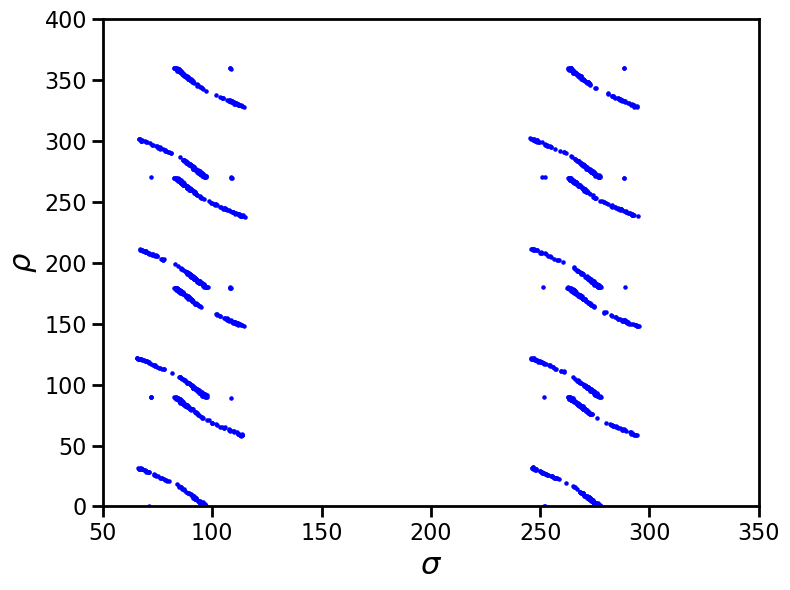}  
        \caption{}
        \label{fig6:a}
    \end{subfigure}\hfill
    \begin{subfigure}[b]{0.45\textwidth}
        \centering
        \includegraphics[width=\textwidth]{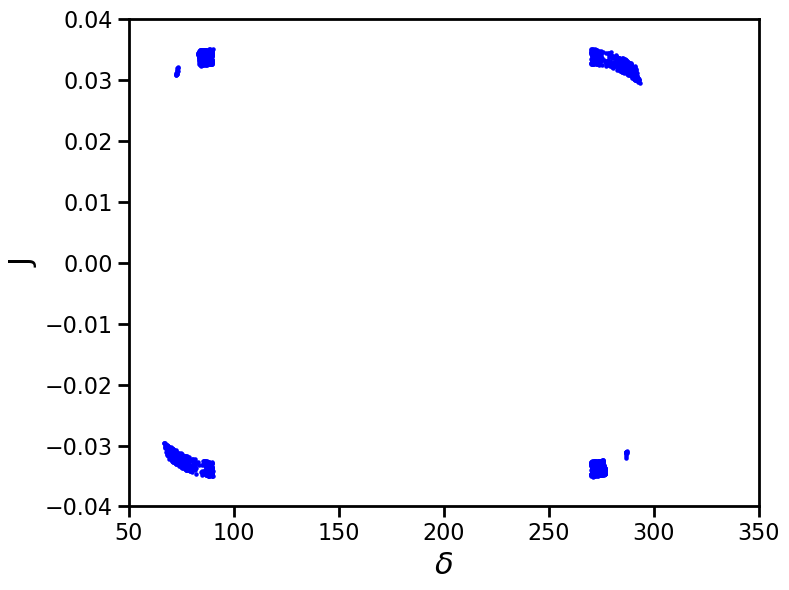}  
        \caption{}
        \label{fig6:b}
    \end{subfigure}
    
\caption{Correlation between Majorana phases $\sigma$ and $\rho$ (a) and Jarsklog Invariant J and $\delta$ (b) for mass matrix $m_{II}$ in NH for $X_1$.}
\label{fig6}
\end{figure}

\begin{figure}[H]  
    \centering
    
    \begin{subfigure}[b]{0.45\textwidth}
        \centering
        \includegraphics[width=\textwidth]{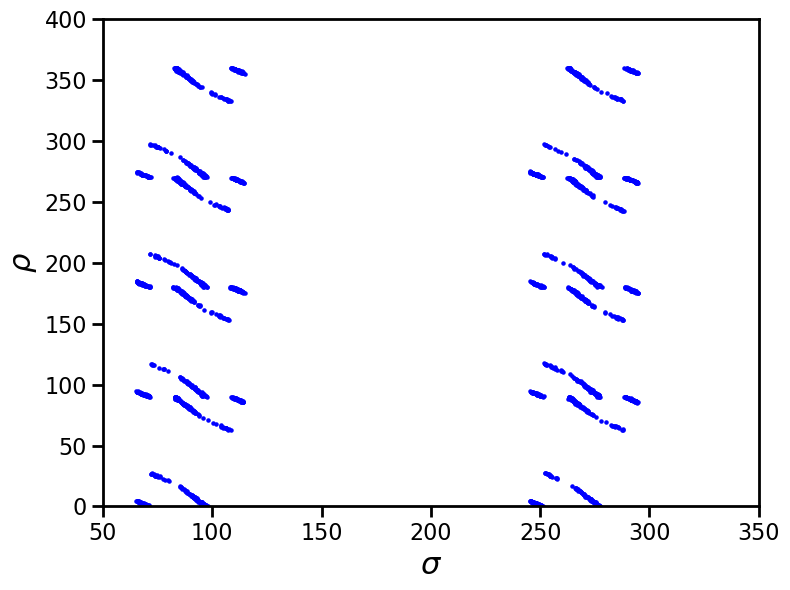}  
        \caption{}
        \label{fig7:a}
    \end{subfigure}\hfill
    \begin{subfigure}[b]{0.45\textwidth}
        \centering
        \includegraphics[width=\textwidth]{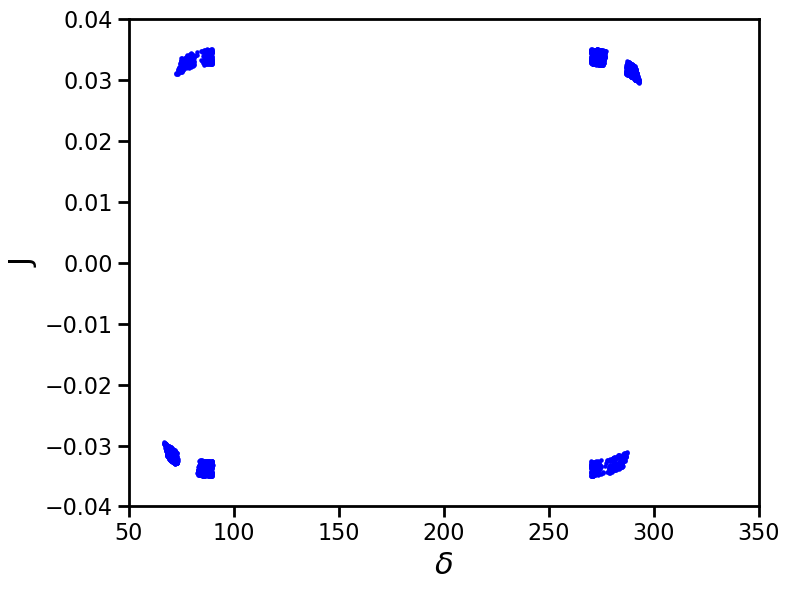}  
        \caption{}
        \label{fig7:b}
    \end{subfigure}
    
\caption{Correlation between Majorana phases $\sigma$ and $\rho$ (a) and Jarsklog Invariant J and $\delta$ (b) for mass matrix $m_{III}$ in NH for $X_1$.}
\label{fig7}
\end{figure}

We have presented the correlation plots of various parameters for matrices \( m_{II} \) and \( m_{III} \), as only they are compatible with the Planck data as well as with P20H+DESI/SDSS+DES-SN dataset.
Figure \ref{fig6}(\subref{fig6:a}), shows the correlation between the two Majorana phases $\rho$ and $\sigma$ for matrix  \( m_{II} \). Figure \ref{fig6}(\subref{fig6:b}) shows the correlation between Jarsklog invariant(J) and Dirac CP phase($\delta$) for matrix  \( m_{II} \), which shows the allowed parameter space. Figure \ref{fig7}(\subref{fig7:a}) also shows the correlation between the two Majorana phases $\rho$ and $\sigma$ for matrix  \( m_{III} \). Figure \ref{fig7}(\subref{fig7:b}) shows the correlation between Jarsklog invariant and Dirac CP phase for matrix  \( m_{III} \).

\begin{figure}[H]  
    \centering
    
    \begin{subfigure}[b]{0.4\textwidth}
        \centering
        \includegraphics[width=\textwidth]{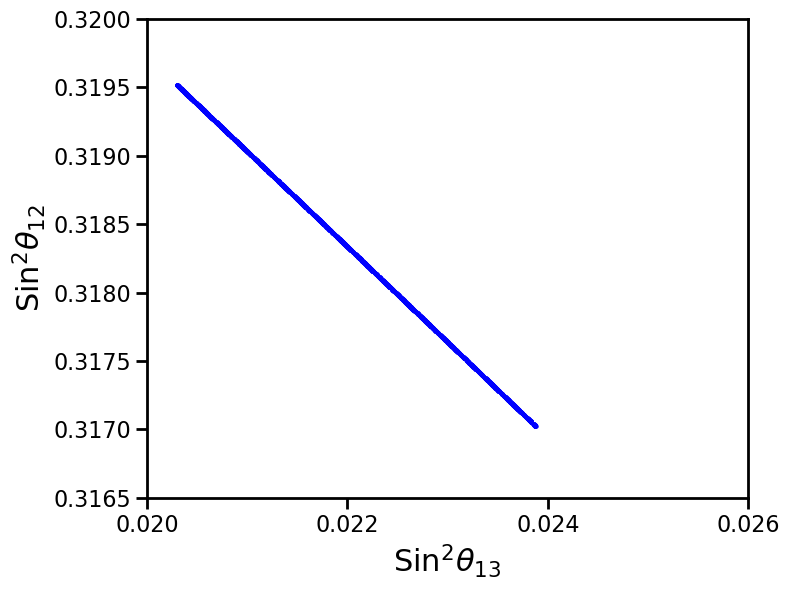}  
        \caption{}
        \label{fig8:a}
    \end{subfigure}\hfill
    \begin{subfigure}[b]{0.45\textwidth}
        \centering
        \includegraphics[width=\textwidth]{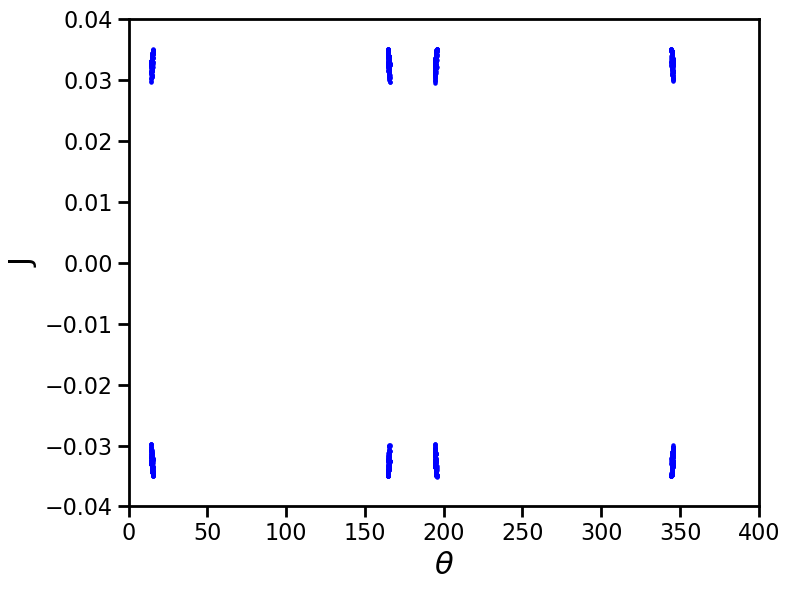}  
        \caption{}
        \label{fig8:b}
    \end{subfigure}
\caption{Correlation between mixing angle $\theta_{23}$ and $\theta_{12}$ for matrix $m_{III}$ in NH (a) and J and rotation angle $\theta$ for matrix $m_{II}$ in NH (b) for $X_1$ condition.}
\label{fig8}
\end{figure}

\begin{figure}[H]  
    \centering
    
    \begin{subfigure}[b]{0.45\textwidth}
        \centering
        \includegraphics[width=\textwidth]{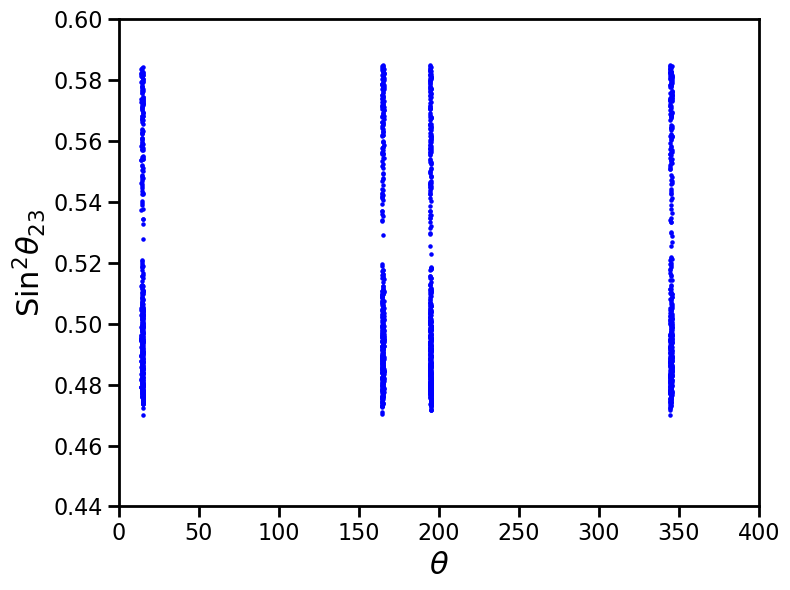}  
        \caption{}
        \label{fig9:a}
    \end{subfigure}\hfill
    \begin{subfigure}[b]{0.45\textwidth}
        \centering
        \includegraphics[width=\textwidth]{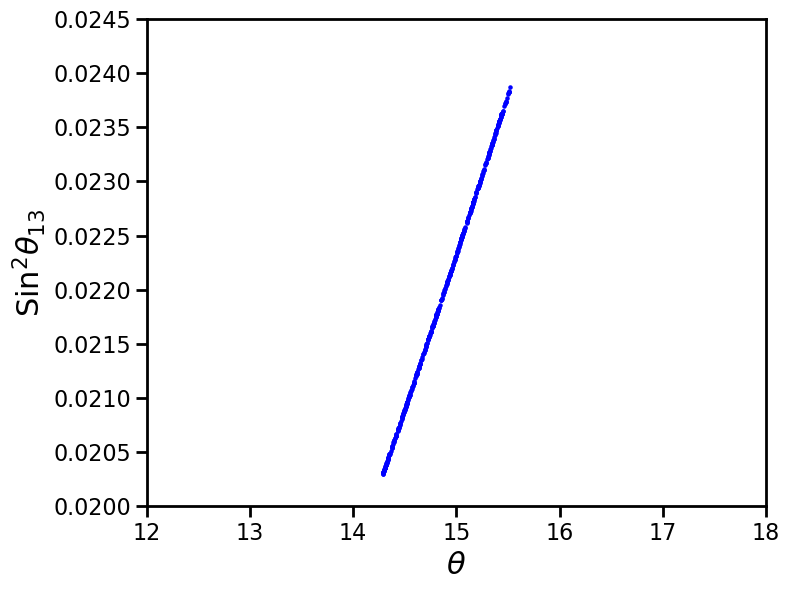} 
        \caption{}
        \label{fig9:b}
    \end{subfigure}
\caption{Correlation between mixing angle $\theta_{23}$ and $\theta$ for matrix $m_{II}$ in NH (a) and $\theta_{13}$ and rotation angle $\theta$ for matrix $m_{III}$ in NH (b) for $X_1$ condition.}
\label{fig9}
\end{figure}

\begin{figure}[H]  
    \centering
    
    \begin{subfigure}[b]{0.48\textwidth}
        \centering
        \includegraphics[width=\textwidth]{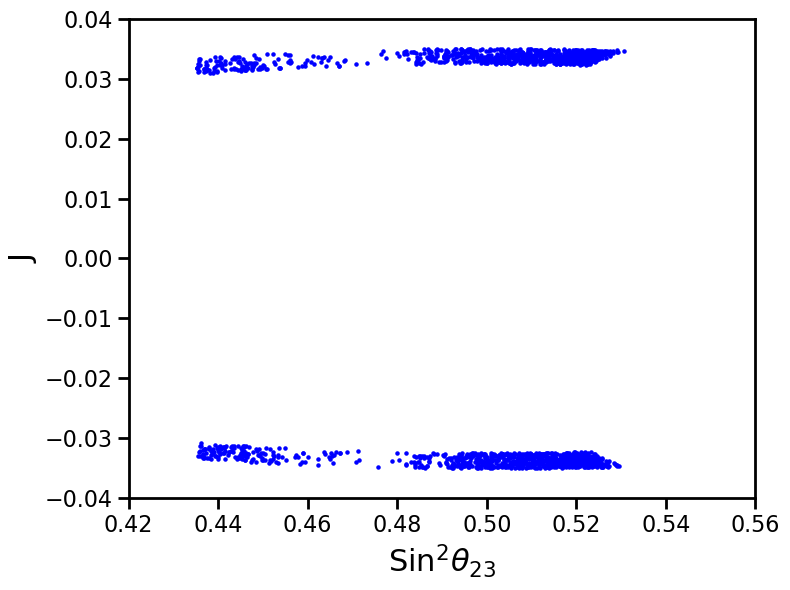}  
        \caption{}
        \label{fig10:a}
    \end{subfigure}\hfill
    \begin{subfigure}[b]{0.48\textwidth}
        \centering
        \includegraphics[width=\textwidth]{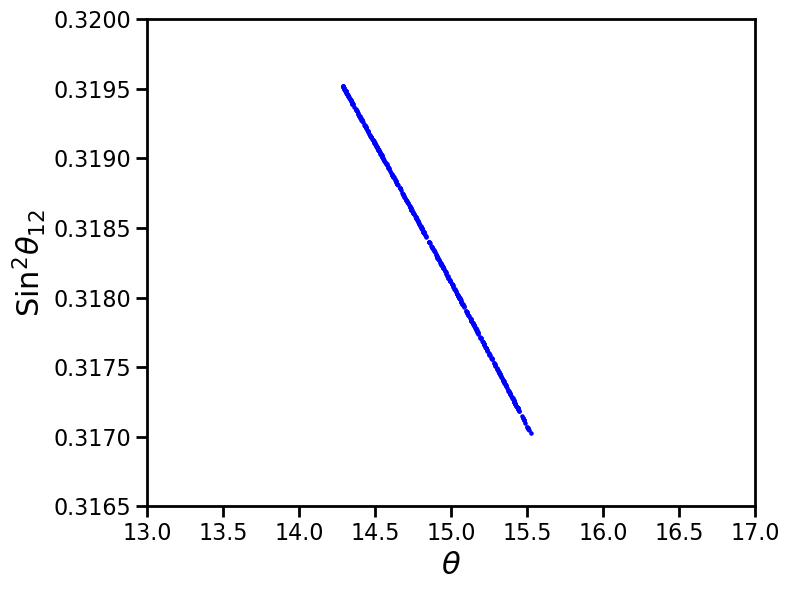}  
        \caption{}
        \label{fig10:b}
    \end{subfigure}

\caption{Correlation between Jarsklog Invariant J and mixing angle $\theta_{23}$ for matrix $m_{III}$ in NH (a) and $\theta_{12}$ and rotation angle $\theta$ for matrix $m_{II}$ in NH (b) for $X_1$ condition.}
\label{fig10}
\end{figure}

Figure \ref{fig8}(\subref{fig8:a}) shows the correlation between the mixing angles $\theta_{12}$ and $\theta_{13}$, constraining the allowed parameter space to $0.3167 < \sin^2\theta_{12} < 0.3195$, and $\sin^2\theta_{13}$ with range ($ 0.020 - 0.024$) for the mass matrix $m_{III}$. Figure \ref{fig8}(\subref{fig8:b}) shows allowed parameter space for J and $\theta$ for mass matrix $m_{II}$. Figure \ref{fig9}(\subref{fig9:a}) shows the correlation between the $\theta_{23}$ and $\theta$, indicates for allowed values for rotation mixing angle within the region \( 0^\circ \leq \theta \leq 11.1^\circ \) and \( 81.1^\circ \leq \theta \leq 88.8^\circ \). Figure \ref{fig9}(\subref{fig9:b}) shows the correlation between the $\theta_{13}$ and $\theta$ for mass matrix $m_{III}$. It is observed that for the $X_1$ condition, $\theta$ is tightly constrained to the range $14.30^\circ - 15.60^\circ$ corresponding to the 3$\sigma$ allowed range of $\theta_{13}$. Figure \ref{fig10}(\subref{fig10:a}) shows the allowed parameter space for J with mixing angle $\theta_{23}$ for mass matrix $m_{III}$. Figure \ref{fig10}(\subref{fig10:b}) shows the allowed parameter space for $\theta$ with 3 $\sigma$ range of $\theta_{12}$ for mass matrix $m_{II}$.

\subsection*{Analysis of $X_2$ Condition}
Similarly for the $X_2$ condition, we have derived \( m_{ee}\) as a function of the lightest neutrino mass, constrained by the $R_\nu$ parameter within the 3$\sigma$ range, aligns with both current and future experimental data as shown in Table \ref{tab3}. 

\begin{table}[H]
  \centering
  \small
  {\begin{tabular*}{\textwidth}{@{\extracolsep{\fill}} c c c c c c}
    \hline
    \hline
    S.No. & Textures & Hierarchy & Effective Majorana mass(eV) & Lightest neutrino mass(eV) & Sum of mass(eV) \\
    \hline
    1 & {$m_I$} & Normal & 0.0 & 0.0021-0.0077 & 0.054-0.071 \\
     &                        & Inverted & - & -&-\\
    \hline
    2 & \multirow{2}{*}{$m_{II}$} & Normal &0.0051-0.30 & 0.0050-0.25&0.065-0.75\\
      &                        & Inverted & 0.023-1.34 & 0.056-0.88&0.20-1.37\\
    \hline
    3 & \multirow{2}{*}{$m_{III}$} & Normal & 0.0050-0.29 & 0.0037-0.24 & 0.063-0.75 \\
      &                        & Inverted & 0.048-0.21 & 0.010-0.15& 0.18-0.75\\
    \hline
    4 & \multirow{2}{*}{$m_{IV}$} & Normal & 0.0068-0.14 &0.051-0.17 & 0.17-0.52 \\
      &                        & Inverted & 0.020-0.21 & 0.052-0.56 & 0.19-0.40\\
    \hline
    5 & \multirow{2}{*}{$m_V$} & Normal & 0.0073-0.20 & 0.050-0.16 & 0.17-0.51\\
      &                        & Inverted &  0.061-0.26&0.058-0.15 & 0.21-0.86\\
    \hline
    6 & \multirow{2}{*}{$m_{VI}$} & Normal & 0.0065-0.10 & 0.051-0.12 & 0.17-0.39 \\
      &                        & Inverted & 0.019-0.17 & 0.052-0.20 & 0.19-0.62 \\
    \hline
    \hline
  \end{tabular*}
  }
  \caption{Values of the effective Majorana mass, lightest neutrino mass, and the sum of the three neutrino masses for both NH and IH for the X$_2$ condition.}

\label{tab3}
\end{table}

The cosmological upper limit on the sum of neutrino masses, \(\Sigma m_i < 0.12\,\text{eV}\), excludes the mass matrices $m_{IV}$, $m_{V}$  and $m_{VI}$ from compatibility. The inverted hierarchy (IH) is excluded from the analysis because all one zero texture matrices in the IH are inconsistent with Planck data and P20$_H$ + DESI/SDSS + DES-SN datasets, within the $\Lambda$CDM + \textit{Fluid DR} + $\sum m_\nu$ model at the 95\% confidence level. The detailed description of textures is given below:

\begin{enumerate}[label=\Roman*.]

    \item The mass matrix \( m_I \), characterized by a zero in the (1,1) position, cannot support inverted hierarchy because it predicts a large value for the reactor mixing angle (\( \theta_{13} \)), which is experimentally not allowed. The imposition of generalized CP conditions does not change this property. For the inverted hierarchy, the neutrino masses satisfy $m_3 < m_1 < m_2$, with typical values $m_1 \approx \sqrt{|\Delta m^2_{\rm atm}|} \sim 0.049~\rm eV$, $m_2 \sim 0.050~\rm eV$, and $m_3 \ll m_1, m_2$. The texture zero condition($m_I = 0$) for $X_2$ is given as
$$
2 m_1 \cos^2\theta + 2 m_3 \sin^2\theta + m_2 e^{-2 i \rho} = 0
$$
can be approximated for estimation by neglecting the small $m_3$ term and taking $e^{-2i\rho} \sim 1$
$$
2 m_1 \cos^2\theta + m_2 \approx 0.
$$
Substituting the representative values gives
$$
2(0.049)\cos^2\theta + 0.050 \approx 0 \quad \Rightarrow \quad \cos^2\theta \approx -0.51,
$$
which is unphysical for real $\theta$.
Physically, this indicates that the small $2 m_3 \sin^2\theta$ term is the only contribution capable of balancing the large positive term $2 m_1 \cos^2\theta + m_2$. Therefore, to satisfy the texture-zero condition, $\sin^2\theta$ must become very large, approaching unity. Mapping this angle to the reactor mixing angle, one finds
$$
\sin^2 2\theta_{13} \gtrsim 0.5,
$$
far exceeding the experimentally allowed value ($\sim 0.085$).
Thus, the texture $m_I$ naturally predicts a large $\theta_{13}$ in the inverted hierarchy, and cannot accommodate this ordering without either fine-tuning the Majorana phase $\rho$, or allowing $m_3$ to be comparable to $m_1, m_2$ (with a relative sign). Similar results have been shown in (\cite{Kumar:2023iaj}). For NH, the effective Majorana mass also vanishes.
    
    \item Figure \ref{fig12}(\subref{fig12:a}) depicts the correlation between the \( m_{ee}\) and \(m_{\text{lightest}}\) for the mass matrix \(m_{II}\) in NH, characterized by a texture zero at the (1,2) position. The analysis reveals that this matrix is consistent with both current and future experimental constraints. The effective Majorana mass spans the range (0.0051 - 0.30)eV, while the lightest neutrino mass varies within (0.0050 - 0.25)eV. Figure \ref{fig12}(\subref{fig12:b}) illustrates the correlation between the effective Majorana mass  and the sum of neutrino masses \(\Sigma m_i\). The results indicate compatibility with the cosmological bound on the sum of neutrino masses, \(\Sigma m_i < 0.12\,\text{eV}\), established by Planck data as well as DESI/SDSS+Pantheon+DES-SN dataset. The sum of neutrino masses is confined to the range (0.065 - 0.75)eV.

\begin{figure}[H]  
    \centering
    
    \begin{subfigure}[b]{0.45\textwidth}
        \centering
        \includegraphics[width=\textwidth]{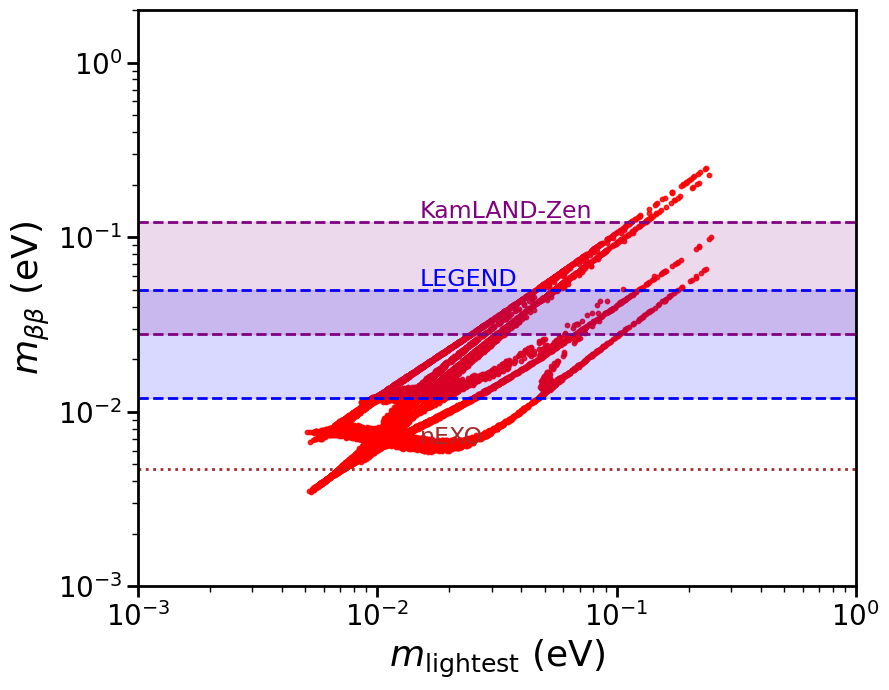}  
        \caption{}
        \label{fig12:a}
    \end{subfigure}\hfill
    \begin{subfigure}[b]{0.45\textwidth}
        \centering
        \includegraphics[width=\textwidth]{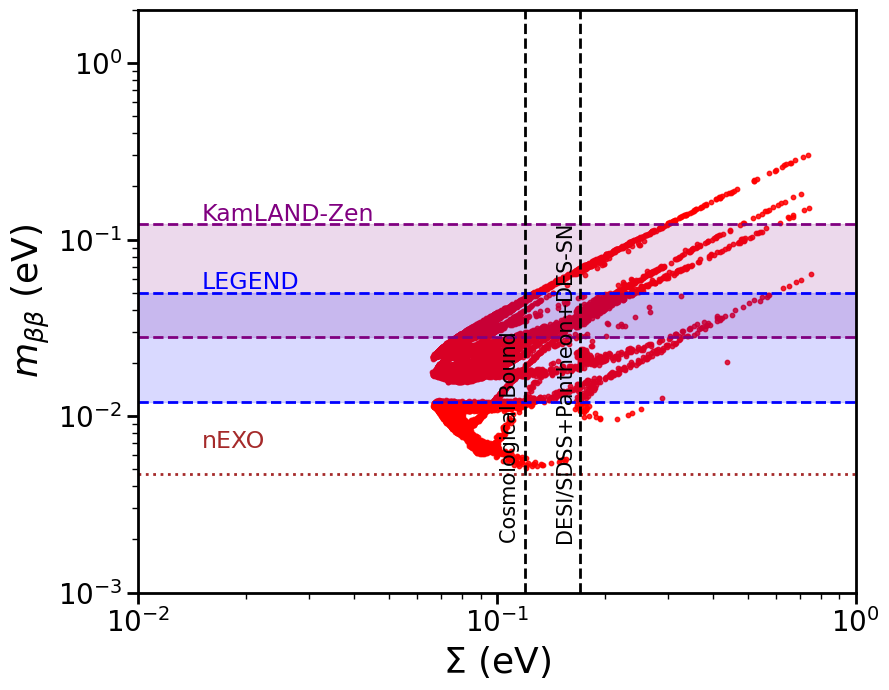}  
        \caption{}
        \label{fig12:b}
    \end{subfigure}

\caption{Effective Majorana mass with respect to lightest Neutrino mass (a) and sum of neutrino mass (b) for mass matrix $m_{II}$ in NH for $X_2$.}
\label{fig12}
\end{figure}

\begin{figure}[H]  
    \centering
    
    \begin{subfigure}[b]{0.45\textwidth}
        \centering
        \includegraphics[width=\textwidth]{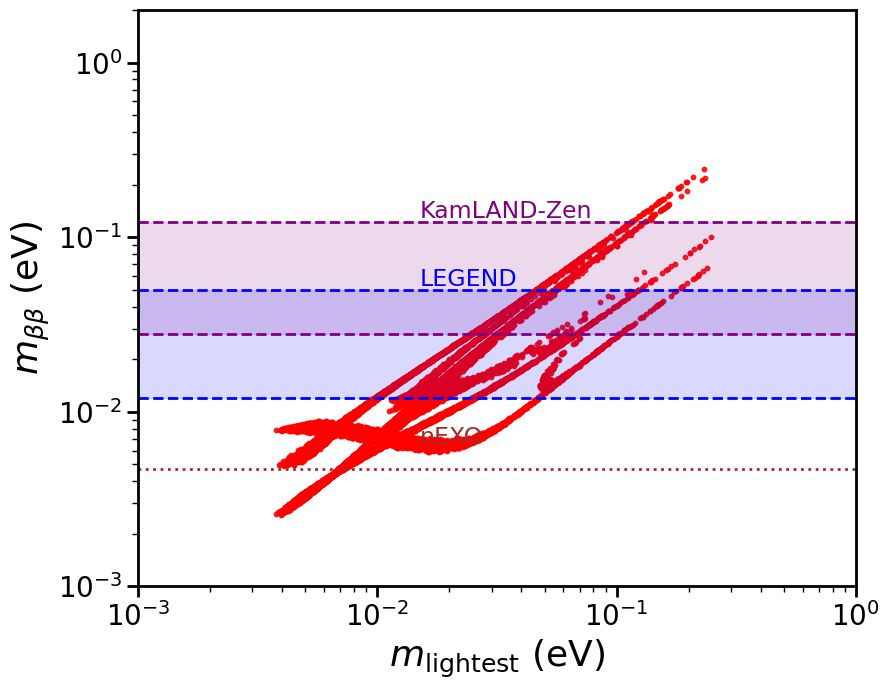}  
        \caption{}
        \label{fig13:a}
    \end{subfigure}\hfill
    \begin{subfigure}[b]{0.45\textwidth}
        \centering
        \includegraphics[width=\textwidth]{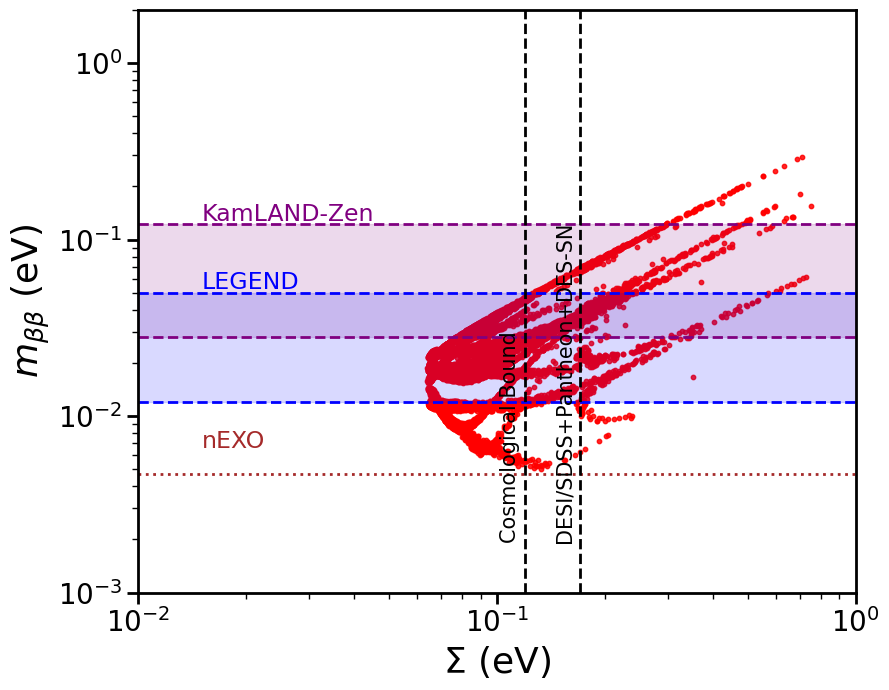}  
        \caption{}
        \label{fig13:b}
    \end{subfigure}

\caption{Effective Majorana mass with respect to lightest neutrino mass (a) and sum of neutrino mass (b) for mass matrix $m_{III}$ in NH for $X_2$.}
\label{fig13}
\end{figure}

    \item Figure \ref{fig13}(\subref{fig13:a}) shows the correlation between the \( m_{ee}\) and \(m_{\text{lightest}}\) for the mass matrix \(m_{III}\), with a zero at the (1,3) position. The analysis shows that this matrix is consistent with both current and future experimental constraints. The effective Majorana mass spans the range (0.0050 - 0.29)eV, while the lightest neutrino mass lies within (0.0037 - 0.24)eV. Figure \ref{fig13}(\subref{fig13:b}) illustrates the correlation between the effective Majorana mass and \(\Sigma m_i\). The results indicate compatibility with the cosmological bound as well as DESI/SDSS+Pantheon+DES-SN dattaset. The sum of neutrino masses is confined to the range (0.063 - 0.75)eV.

    \item Figure \ref{fig14}(\subref{fig14:a}) illustrates the correlation between the \( m_{ee}\) and \(m_{\text{lightest}}\) for the mass matrix \(m_{IV}\), with a zero at the (2,2) position. The analysis shows that, \( m_{ee}\) have lower bound at 0.0068eV, shows compatibility with the KamlandZen and LEGEND experiments. Figure \ref{fig14}(\subref{fig14:b}) shows the correlation between the effective Majorana mass and \(\Sigma m_i\). The results indicate incompatibility with the cosmological bound but compatible with the DESI/SDSS+Pantheon+DES-SN dataset. The sum of neutrino masses is confined to the range (0.17-0.52)eV.

\begin{figure}[H]  
    \centering
    
    \begin{subfigure}[b]{0.48\textwidth}
        \centering
        \includegraphics[width=\textwidth]{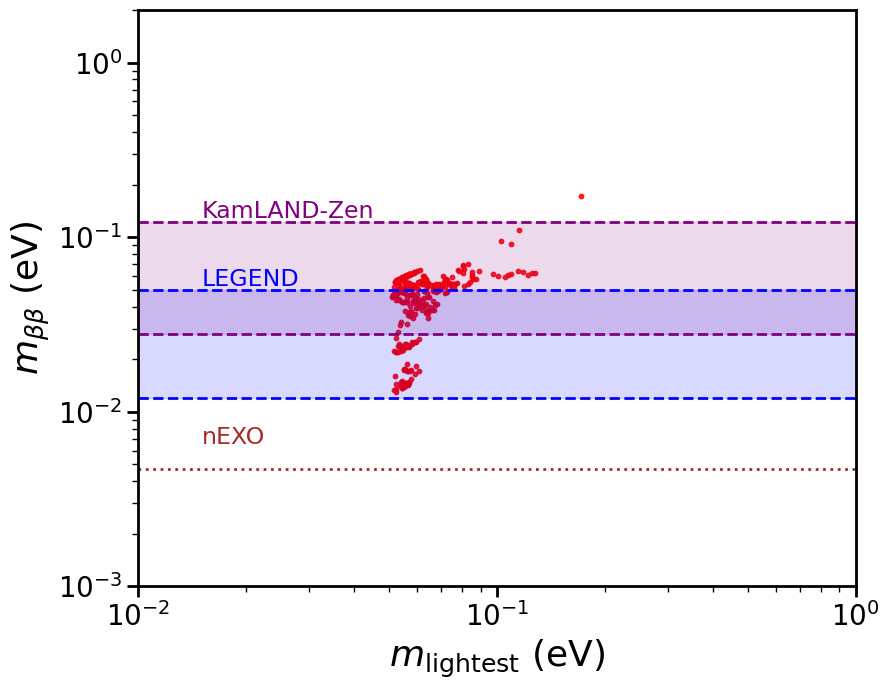}  
        \caption{}
        \label{fig14:a}
    \end{subfigure}\hfill
    \begin{subfigure}[b]{0.48\textwidth}
        \centering
        \includegraphics[width=\textwidth]{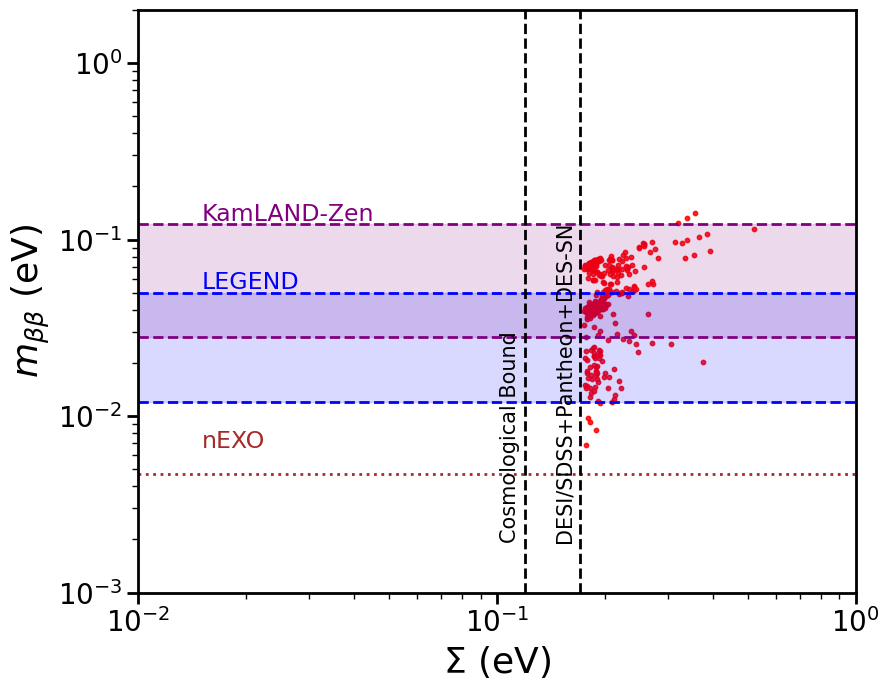}  
        \caption{}
        \label{fig14:b}
    \end{subfigure}

\caption{Effective Majorana mass with respect to lightest neutrino mass (a) and sum of neutrino mass (b) for mass matrix $m_{IV}$ in NH for $X_2$.}
\label{fig14}
\end{figure}

\begin{figure}[H]  
    \centering
    
    \begin{subfigure}[b]{0.48\textwidth}
        \centering
        \includegraphics[width=\textwidth]{23_tm1.png}  
        \caption{}
        \label{fig15:a}
    \end{subfigure}\hfill
    \begin{subfigure}[b]{0.48\textwidth}
        \centering
        \includegraphics[width=\textwidth]{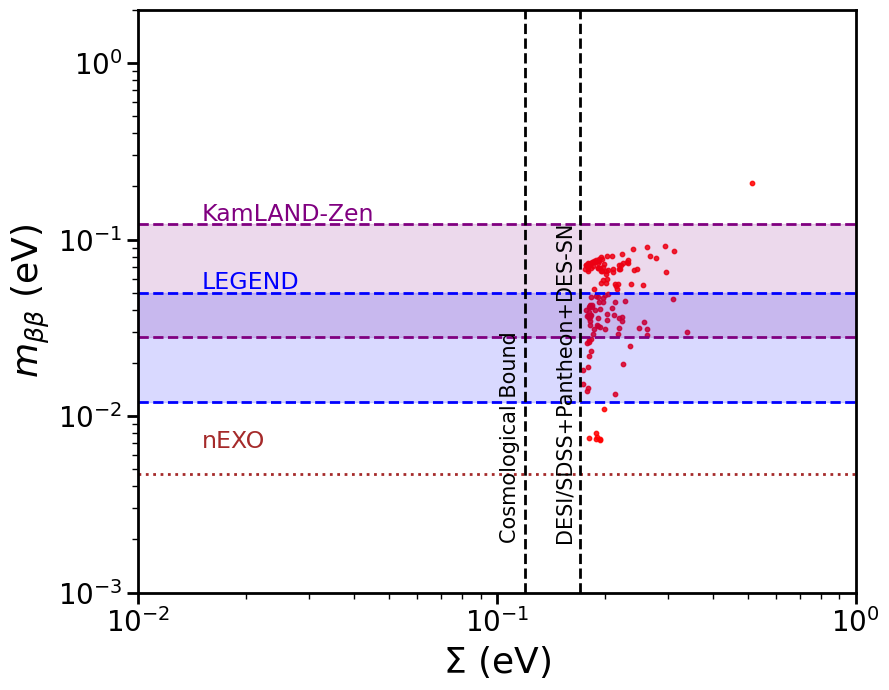}  
        \caption{}
        \label{fig15:b}
    \end{subfigure}

\caption{Effective Majorana mass with respect to lightest neutrino mass (a) and sum of neutrino mass (b) for mass matrix $m_{V}$ in NH for $X_2$.}
\label{fig15}
\end{figure}

\begin{figure}[H]  
    \centering
    
    \begin{subfigure}[b]{0.48\textwidth}
        \centering
        \includegraphics[width=\textwidth]{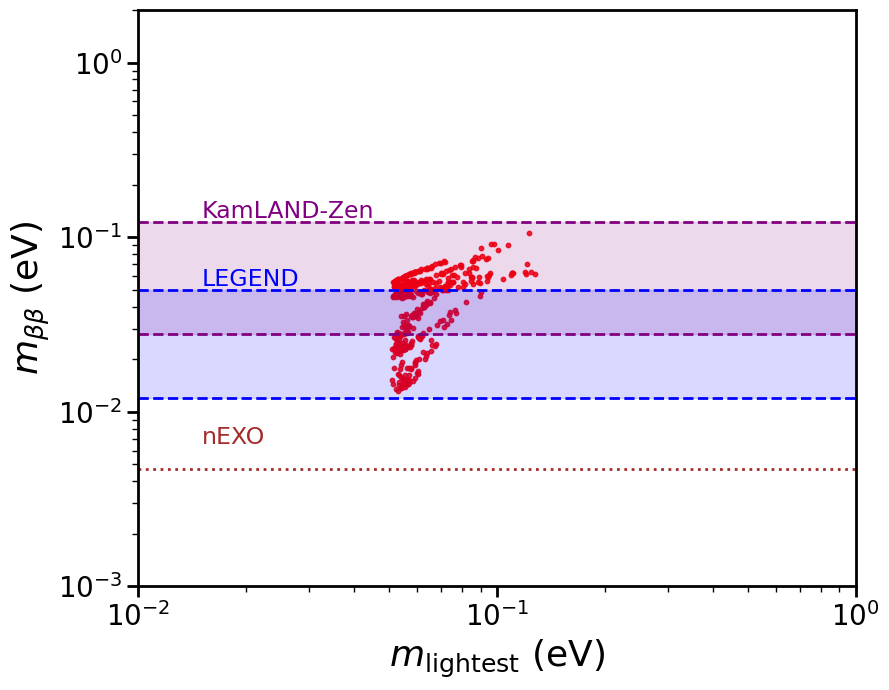}  
        \caption{}
        \label{fig16:a}
    \end{subfigure}\hfill
    \begin{subfigure}[b]{0.48\textwidth}
        \centering
        \includegraphics[width=\textwidth]{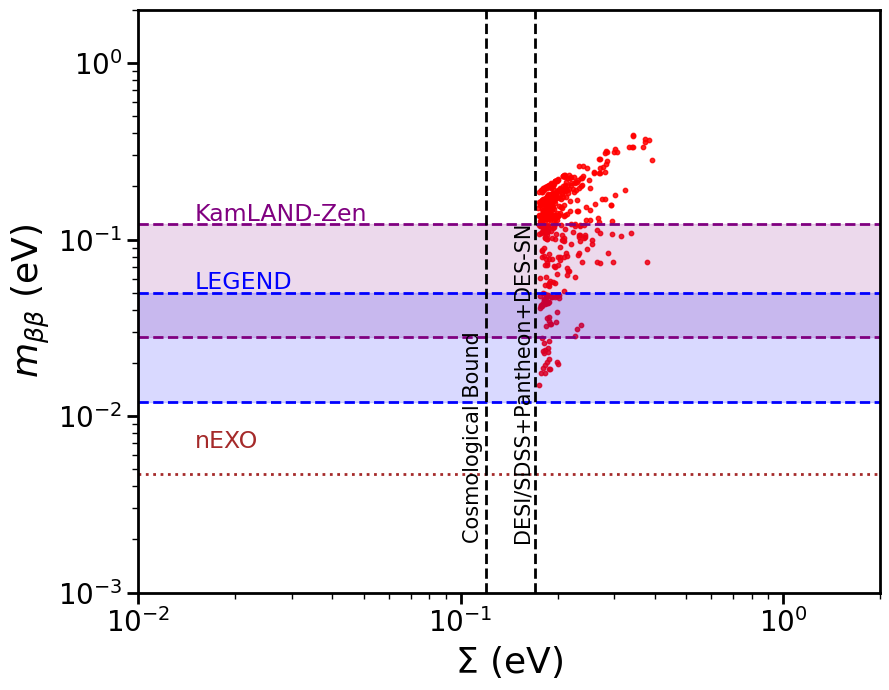}  
        \caption{}
        \label{fig16:b}
    \end{subfigure}

\caption{Effective Majorana mass with respect to lightest neutrino mass (a) and sum of neutrino mass (b) for mass matrix $m_{VI}$ in NH for $X_2$.}
\label{fig16}
\end{figure}

    \item Figure \ref{fig15}(\subref{fig15:a}) depicts the correlation between the \( m_{ee}\) and \(m_{\text{lightest}}\) for the mass matrix \(m_{V}\), having zero at the (2,3) position. The analysis provides lower bound on \( m_{ee}\), shows compatibility with the KamlandZen and LEGEND experiment.  Figure \ref{fig15}(\subref{fig15:b}) illustrates the correlation between the \( m_{ee}\) and \(\Sigma m_i\). The results indicate in compatibility with the cosmological bound but compatible with DESI/SDSS+Pantheon+DES-SN dataset. The sum of neutrino masses is confined to the range (0.17-0.51)eV.

    \item Figure \ref{fig16}(\subref{fig16:a}) shows the correlation between the \( m_{ee}\) and the \(m_{\text{lightest}}\) for the mass matrix \(m_{VI}\), characterized by a zero at the (3,3) position. The analysis shows that this matrix is consistent with current experimental constraints. The effective Majorana mass spans the range (0.0065 - 0.10)eV, while the lightest neutrino mass lies within (0.051 - 0.12)eV. Figure \ref{fig16}(\subref{fig16:b}) depicts the correlation between the \( m_{ee}\) and the  \(\Sigma m_i\). The results indicate incompatibility with the cosmological bound but compatible with the DESI/SDSS+Pantheon+DES-SN datatset. The sum of neutrino masses is confined to the range (0.17 - 0.39)eV. 
    
\end{enumerate}

\begin{figure}[H]  
    \centering
    
    \begin{subfigure}[b]{0.48\textwidth}
        \centering
        \includegraphics[width=\textwidth]{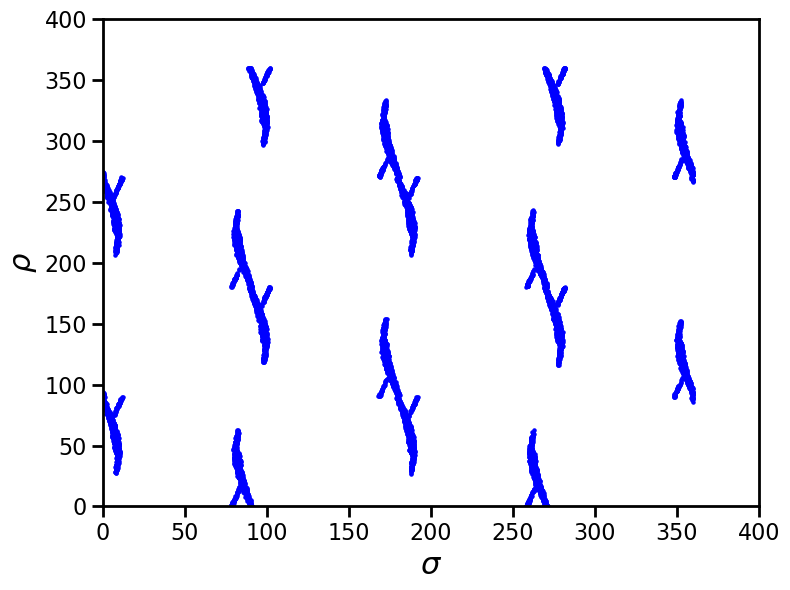}  
        \caption{}
        \label{fig17:a}
    \end{subfigure}\hfill
    \begin{subfigure}[b]{0.48\textwidth}
        \centering
        \includegraphics[width=\textwidth]{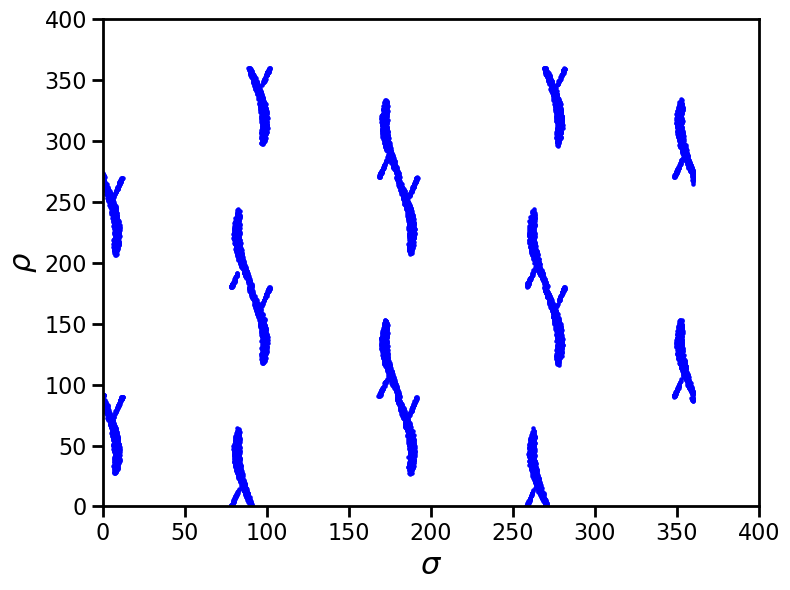}  
        \caption{}
        \label{fig17:b}
    \end{subfigure}

\caption{Correlation between Majorana phases $\sigma$ and $\rho$ for mass matrix $m_{II}$ in NH (a) and Majorana phases $\sigma$ and $\rho$ for mass matrix $m_{III}$ in NH (b) for $X_2$. }
\label{fig17}
\end{figure}

\begin{figure}[H]  
    \centering
    
    \begin{subfigure}[b]{0.45\textwidth}
        \centering
        \includegraphics[width=\textwidth]{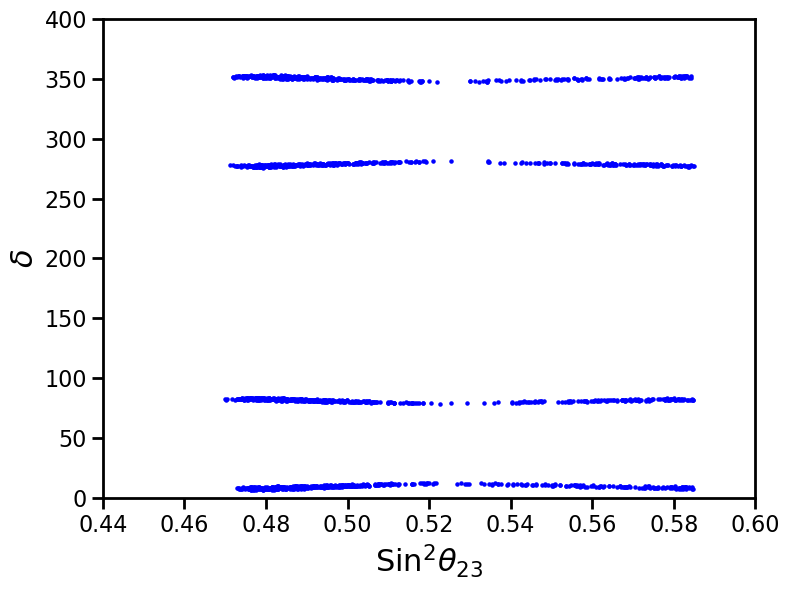}  
        \caption{}
        \label{fig18:a}
    \end{subfigure}\hfill
    \begin{subfigure}[b]{0.45\textwidth}
        \centering
        \includegraphics[width=\textwidth]{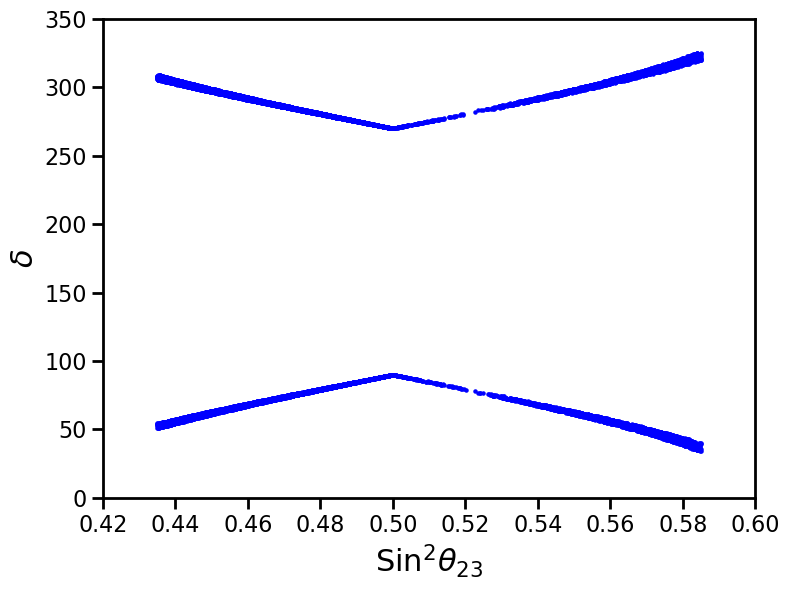}  
        \caption{}
        \label{fig18:b}
    \end{subfigure}
\caption{Correlation between mixing angle $\theta_{23}$ and $\delta$ for matrix $m_{II}$ in NH for $X_1$ condition (a) and J and $\delta$ for matrix $m_{II}$ in NH (b) for $X_2$ condition.}
\label{fig18}
\end{figure}

The Figure \ref{fig17}(\subref{fig17:a}) and \ref{fig17}(\subref{fig17:b}) shows the allowed parameter space for matrices $m_{II}$ and $m_{III}$, respectively, which indicates that $\rho$ is sharply constrained in the region around $\sigma = 0, \frac{\pi}{2}, \pi, \frac{3\pi}{2}, 2\pi$. The different mixing matrix, as shown in Eqn. \ref{eqn8} and \ref{eqn10}, predicts specific correlations between the mixing angle $\theta_{23}$ and the CP-violating phase $\delta$ when one zero textures are introduced. As $\theta_{23}$ tends towards maximal mixing, the corresponding CP violation is also expected to reach maximality, with $\delta$ approaching values of $\pi/2$ or $3\pi/2$ shown in Figure \ref{fig18}(\subref{fig18:a}) and \ref{fig18}(\subref{fig18:b}).

We also demonstrate correlation between the mixing angles $\theta_{12}$, $\theta_{23}$, and the free parameter $\theta$. 

\begin{figure}[H]  
    \centering
    
    \begin{subfigure}[b]{0.5\textwidth}
        \centering
        \includegraphics[width=\textwidth]{j_theta_tm1_m12.png}  
        \caption{}
        \label{fig19:a}
    \end{subfigure}\hfill
    \begin{subfigure}[b]{0.45\textwidth}
        \centering
        \includegraphics[width=\textwidth]{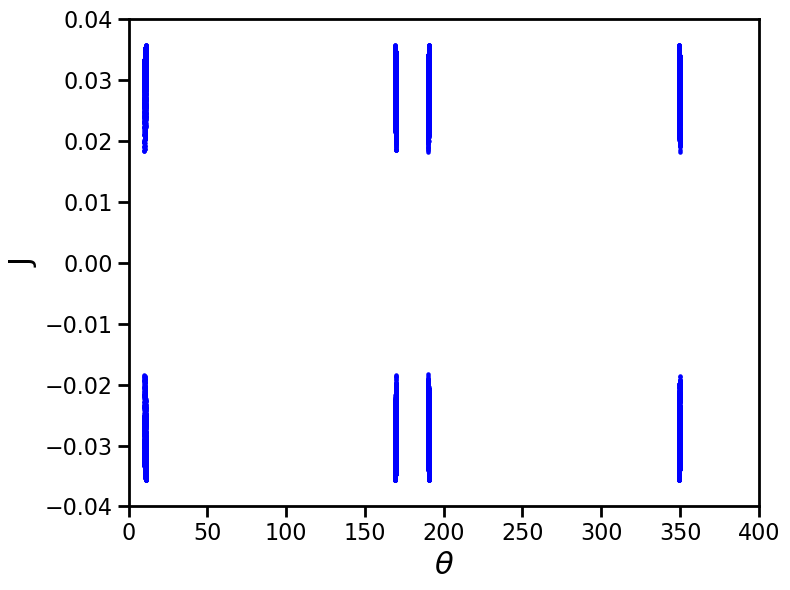}  
        \caption{}
        \label{fig19:b}
    \end{subfigure}
\caption{Correlation between rotation angle $\theta$ and J for matrix $m_{II}$ in NH (a) and $\theta_{12}$ and $\theta_{23}$ for matrix $m_{III}$ in NH (b) for $X_2$ condition.}
\label{fig19}
\end{figure}

\begin{figure}[H]  
    \centering
    
    \begin{subfigure}[b]{0.45\textwidth}
        \centering
        \includegraphics[width=\textwidth]{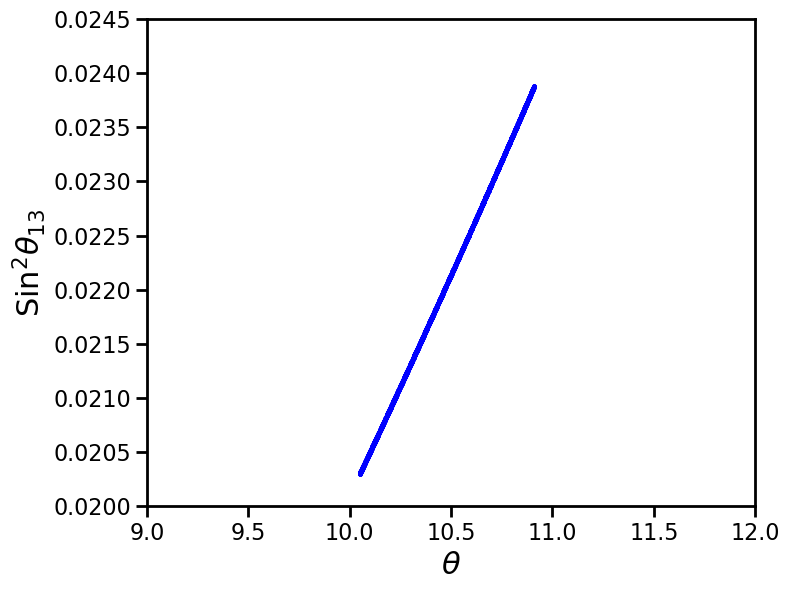}  
        \caption{}
        \label{fig20:a}
    \end{subfigure}\hfill
    \begin{subfigure}[b]{0.45\textwidth}
        \centering
        \includegraphics[width=\textwidth]{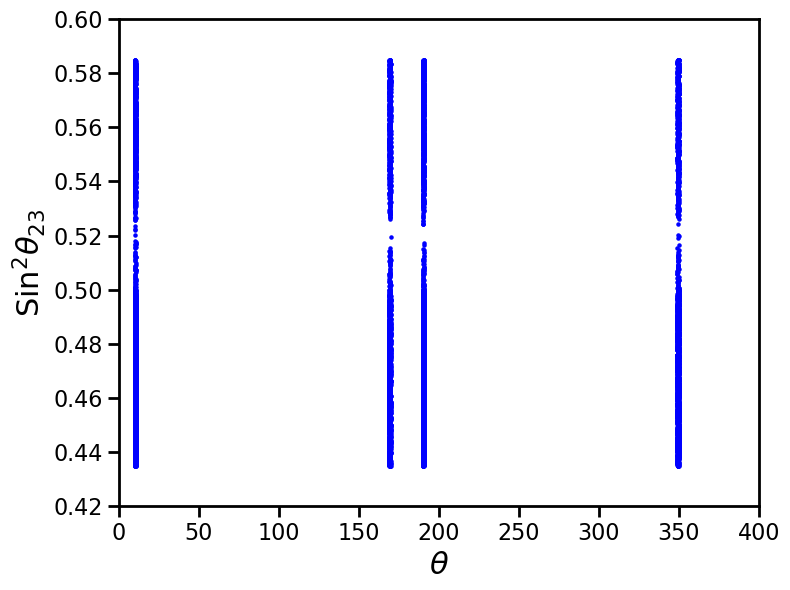} 
        \caption{}
        \label{fig20:b}
    \end{subfigure}
    
\caption{Correlation between mixing angle $\theta_{13}$ and $\theta$ (a) and $\theta_{23}$ and $\theta$ (b) for $X_2$ for mass matrix $m_{III}$ in NH.}
\label{fig20}
\end{figure}

\begin{figure}[H]  
    \centering
    
    \begin{subfigure}[b]{0.48\textwidth}
        \centering
        \includegraphics[width=\textwidth]{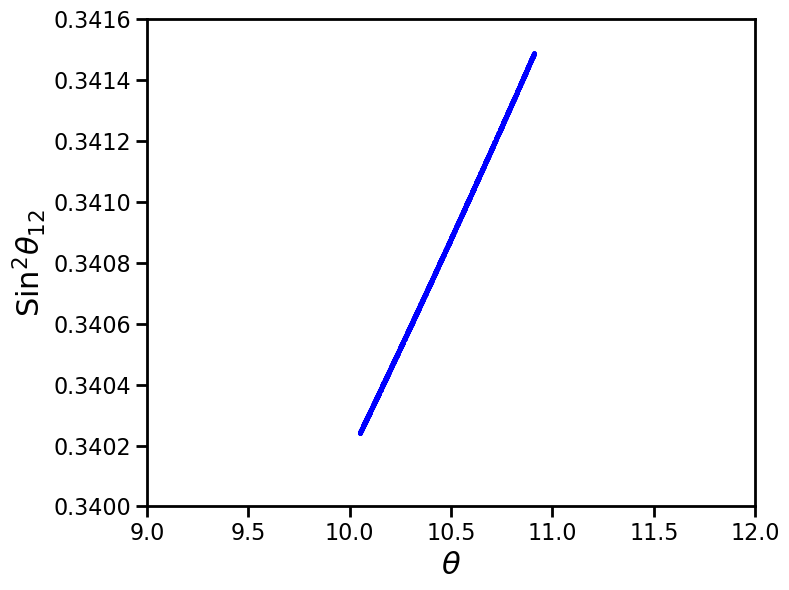}  
        \caption{}
        \label{fig21:a}
    \end{subfigure}\hfill
    \begin{subfigure}[b]{0.48\textwidth}
        \centering
        \includegraphics[width=\textwidth]{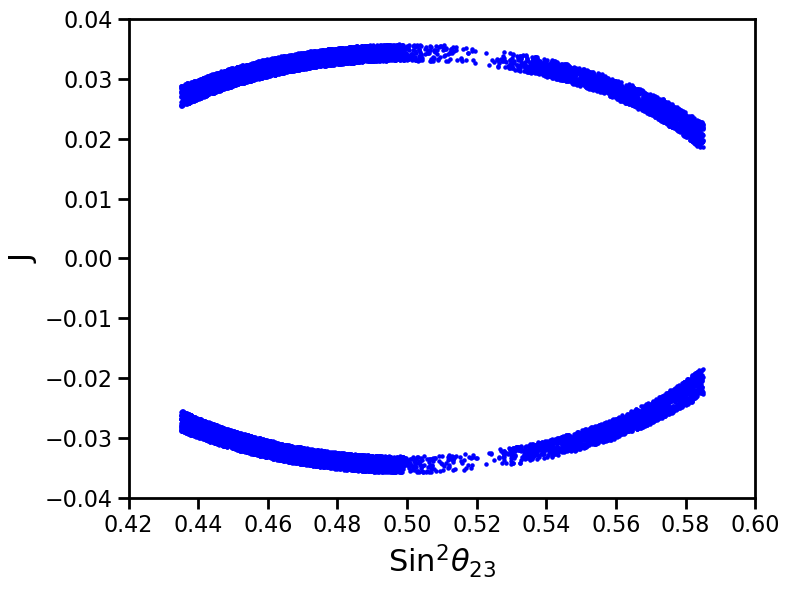}  
        \caption{}
        \label{fig21:b}
    \end{subfigure}

\caption{Correlation between $\theta_{12}$ and $\theta$ (a) and Jarsklog Invariant J and mixing angle $\theta_{23}$ (b) for matrix $m_{II}$ in NH for $X_2$ condition .}
\label{fig21}
\end{figure}

Figure \ref{fig19}(\subref{fig19:a}) depicts allowed parameter space for J with respect to the rotation angle $\theta$ for mass matrix $m_{II}$. The correlation between the mixing angles \(\theta_{12}\) and \(\theta_{23}\) for mass matrix $m_{III}$, shown in Figure \ref{fig19}(\subref{fig19:b}), indicates the highly constrained region for \(\theta_{12}\) and \(\theta_{23}\). The allowed region for \(\theta\) with respect to 3$\sigma$ ranges of the mixing angles \(\theta_{13}\) and \(\theta_{23}\) are presented in the Figures \ref{fig20}(\subref{fig20:a}) and \ref{fig20}(\subref{fig20:b}) for mass matrix $m_{III}$. The correlation between \(\theta\) and \(\theta_{12}\) for mass matrix $m_{II}$, is shown in \ref{fig21}(\subref{fig21:b}). The Figure \ref{fig21}(\subref{fig21:b}) shows the allowed region of parameter J with respect to the mixing angle \(\theta_{23}\) for mass matrix $m_{II}$.

\section{Conclusions}

In this paper, we have analyzed the implications of generalized CP symmetries and one zero textures on neutrino mass matrices. We found that the presence of one zero textures, when combined with generalized CP symmetries such as $X_1$ and $X_2$, leads to specific and predictive patterns for the neutrino mass matrix. To obtain the generalized CP conditions, we employed the cTBM matrix, which further constrain the Majorana phases. The presence of one zero textures leads to six distinct patterns for the neutrino mass matrix. For matrices with one zero textures, all cases are allowed except for $m_{I} = 0$ in the inverted hierarchy because this condition leads to an excessively large value of reactor mixing angle $\theta_{13}$. The presence of the generalized CP conditions in the neutrino mass  matrix does not change this property.

We computed the effective Majorana mass as a function of the lightest neutrino mass and the sum of neutrino masses. The resulting effective Majorana mass is consistent with the current limits from the KamlandZen and LEGEND experiments. We have calculated the value of the lightest neutrino mass within the constrained region defined by \( R_\nu \) in the \( 3\sigma \) range. Additionally, we computed the sum of the three neutrino masses, and by applying the cosmological constraints from Planck data and DESI/SDSS+Pantheon+DES-SN dataset, we found that the inverted mass hierarchy is disfavored by this data. The results for all the matrices are summarized in Tables \ref{Tab2} and \ref{tab3}. The DESI experiment has set a stringent bound on the sum of neutrino masses ($0.072$ eV) at 95\% CL \cite{Allali:2024aiv}. In the $X_2$ case, the mass matrix $m_I$, $m_{II}$ and $m_{III}$ respect this bound, while the rest are incompatible with it. In the future, this bound may rule out all the matrices with $X_1$ case except $m_I$ and all other matrices in the $X_2$ scenario. Our analysis also shows distinct correlations between the mixing angles and the Dirac CP phase \( \delta \) for the CP symmetries \( X_1 \) and \( X_2 \). These correlations serve as identifying features of each texture. In the correlation, we examined the different mixing angles and rotation angle \( \theta \) within their \( 3\sigma \) ranges. The resulting correlations illustrate a very constrained region of \( \theta \). We also present the plots between Jarlskog invariant \( J \) and the mixing angles, which exhibit sharp correlations for both conditions, \( X_1 \) and \( X_2 \). Similarly, we investigated the effective Majorana mass as a function of \( \delta \), which shows compatibility with current experimental data. From the correlation, we find that the rotation angle $\theta$ is constrained between ($14.3^\circ - 15.6 ^\circ$) for the $X_1$ condition, and between ($10.1^\circ - 11.1^\circ$) for the $X_2$ condition. 

In summary, one zero textures combined with generalized CP symmetries provide predictive and testable structures for the neutrino mass matrix. 
The predictions for the atmospheric mixing angle $\theta_{23}$ can be tested at NO$\nu$A \cite{Frank:2023rwq}, DUNE\cite{DUNE:2020fgq}, and Hyper-Kamiokande \cite{Hyper-Kamiokande:2018ofw} experiments. The predicted effective Majorana mass is consistent with current limits from KamlandZen and LEGEND and remains within the reach of future neutrinoless double beta decay experiments. This framework can be tested with upcoming results from oscillation and neutrinoless double beta decay experiments.

\section*{Acknowledgements}
The authors are thankful to the Inter University Centre for Astronomy and Astrophysics (IUCAA), Pune for providing necessary facilities during the completion of this work.

\end{document}